\documentclass[journal=jctcce,manuscript=article,layout=traditional]{achemso}

\usepackage{braket}
\usepackage{bm}
\usepackage{upgreek}
\newcommand{\mr}{\mathrm}
\newcommand{\phansup}{^{\vphantom{\mathrm{d}}}}

\usepackage[version=3]{mhchem} 

\usepackage{array}
\usepackage{hhline}
\usepackage{colortbl}
\usepackage{multirow} 
\usepackage{longtable}
\usepackage{makecell}
\usepackage[usenames,dvipsnames,table]{xcolor}

\usepackage{algorithm2e, setspace}

\usepackage[authormarkuptext=name,authormarkup=none,final]{changes}
\definecolor{acolour}{RGB}{153, 76, 0}
\definecolor{bcolour}{RGB}{255, 0, 0}
\definechangesauthor[name={Aleksei}, color={acolour}]{AVI}
\definechangesauthor[name={Gianluca}, color={bcolour}]{GL}
\definechangesauthor[name={Rev1}, color=BrickRed]{Rev1}
\definechangesauthor[name={Rev2}, color=NavyBlue]{Rev2}
\definechangesauthor[name={Other}, color=YellowOrange]{Other}
\usepackage{todonotes}

\usepackage{float}

\usepackage{hyperref}

\usepackage{appendix}

\newcolumntype{M}[1]{ >{\centering\arraybackslash} m{#1} }


\author{Gianluca Levi}
\affiliation[University Iceland]
{Science Institute and Faculty of Physical Sciences, University of Iceland, 107
Reykjav\'{i}k, Iceland}
\email{giale@hi.is}
\author{Aleksei V. Ivanov}
\affiliation[University Iceland]
{Science Institute and Faculty of Physical Sciences, University of Iceland, 107
Reykjav\'{i}k, Iceland}
\alsoaffiliation[University St Petersburg]
{Saint Petersburg State University, 199034 Saint Petersburg, Russia.}
\author{Hannes J\'{o}nsson}
\affiliation[University Iceland]
{Science Institute and Faculty of Physical Sciences, University of Iceland, 107
Reykjav\'{i}k, Iceland}

\title[do-mom]
{Variational density functional calculations of excited states via direct optimization} 

\abbreviations{}

\begin{document}

\renewcommand*\tocentryname{TOC Graphic}
\begin{tocentry}
   \includegraphics{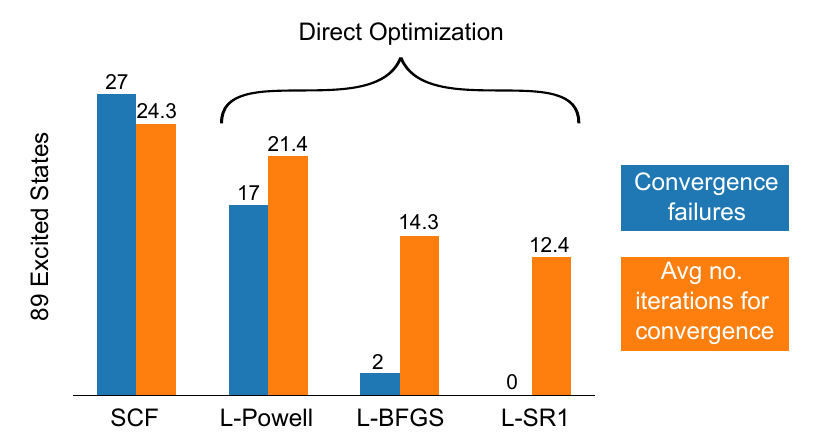}
\end{tocentry}

\begin{abstract}
The development of variational density functional theory approaches to excited electronic states
is impeded by limitations of the commonly used self-consistent field (SCF) procedure.
A method based on a direct optimization approach as well as the maximum overlap method 
is presented and the performance compared with previously proposed SCF strategies.
Excited-state solutions correspond to saddle points of the energy as a function of the electronic 
degrees of freedom. The approach presented here makes use of a preconditioner determined with the help of 
the maximum overlap method to guide the convergence on a target $n$th-order saddle point. 
The method is found to be more robust and to converge faster than previously proposed SCF approaches
for a set of \added[id=Rev2]{89} excited states of molecules. 
A limited-memory formulation of the symmetric rank-one method for updating the inverse Hessian 
is found to give the best performance. 
A conical intersection for the carbon monoxide molecule is calculated without resorting to fractional occupation numbers. 
\added[id=Rev2]{Calculations on excited states of the hydrogen atom and a doubly excited state of the dihydrogen 
molecule using a self-interaction corrected functional are presented. 
For these systems, the self-interaction correction is found to improve the accuracy 
of density functional calculations of excited states.}\\
\end{abstract}

\section{Introduction}
In the light of recent and rapid advancements in fields such as photocatalysis 
and ultrafast spectroscopies, 
the availability of efficient and accurate computational methods to model 
electronic excited-state properties of molecules has become increasingly important.
A widely used methodology to obtain excited-state properties of molecules is 
time-dependent density functional theory (TDDFT)\cite{Dreuw2005, Casida1995, Runge1984}.
Practical applications of TDDFT rely on (i) linear response to describe the perturbation 
of the electron density due to an external field, and (ii) the adiabatic approximation, 
which neglects the time dependency of the functional derivative of the exchange-correlation 
(xc) potential with respect to the density, the so-called xc kernel. With those approximations, 
the computations can be carried out with local and semi-local ground-state 
Kohn-Sham (KS)\cite{Kohn:1965, Hohenberg:1964} functionals without excessive computational requirements and
this has been found to give an adequate description 
of valence excitations in many cases\cite{VanMeer2014, Dreuw2005}. 
On the other hand, the neglect of the time dependency of the xc kernel limits the applicability 
of this approach and makes it, for example, inadequate for the description of double 
excitations\cite{Levine2006, Maitra2004, Tozer2000} and 
conical intersections between ground and excited states\cite{Huix-Rotllant2016, Levine2006}.
Moreover, due to the incorrect form of the potential at long range and to the lack of 
orbital relaxation effects\cite{Zhao2019, Zhekova2014, Ziegler2008}, TDDFT with KS functionals 
suffers from systematic errors when applied to excited states that are diffuse, such as 
Rydberg states\cite{Seidu2015, VanMeer2014, Cheng2008}, or involve transfer of charge between 
spatially separated regions\cite{Baerends2013,Dreuw2004, Dreuw2003}.

Some of these issues can be solved employing alternative DFT formulations 
where excited states are obtained as single Slater determinant wave 
functions optimized for non-aufbau occupations using ground-state functionals. 
Here, one seeks a saddle point on the energy surface instead of a minimum.
Thanks to the inclusion of state-specific orbital 
relaxation effects, these methods can describe a broader range of excited states than linear-response 
TDDFT in the adiabatic approximation, and have, therefore, seen a revival of interest recently
\cite{Malis2020, Hait2020b, Hait2020, Pradhan2018, Barca2018, Levi2018, Pavanello2018, Liu2017, Park2016, Seidu2015, 
Park2015, Zhekova2014, Peng2013, Maurer2013, Himmetoglu2012, Kowalczyk2011, Maurer2011,
Gilbert2008, gavnholt:2008, Cheng2008}.

The excited-state DFT methodology that we consider here does not enforce orthogonality constraints between the
different excited-state solutions and the ground state, and, therefore, represents a straightforward 
extension of ground-state DFT\footnote[4]
{Sometimes, this method is \added[id=Other]{referred to} as $\Delta$ self-consistent field ($\Delta$SCF)
\cite{Hait2020, Levi2018, Park2015, Maurer2011, Kowalczyk2011, gavnholt:2008},
but here we prefer the more general term ``excited-state DFT'', following Cheng \emph{et al.}\cite{Cheng2008}, 
avoiding the risk of relating the method to a specific nonlinear variational procedure (such as SCF) 
and to the computation of a specific excited-state property (the excitation energy through the energy difference, $\Delta$,
between excited and ground state).}. 
Higher-energy stationary points of ground-state 
density functionals obtained in this way do not necessarily represent rigorous approximations 
to the exact stationary states \cite{Cheng2008, Perdew1985}. 
On the other hand, practice has shown that excited-state DFT 
calculations are usually able to deliver useful approximations to excited-state properties
of molecules, such as excitation energies and potential energy surfaces\cite{Barca2018, Gilbert2008}. 
Some studies have also highlighted how the method can satisfactorily treat cases, such as conical 
intersections, with strong static correlation, despite the single-determinant approximation
\added[id=Other]{\cite{Malis2020, Pradhan2018, Pradhan2018, Maurer2011}}.

From a more practical point of view, the lack of orthogonality and the single-determinant 
approximation give rise to difficulties in the convergence of higher-energy solutions.  
First of all, when lower-energy states of the same symmetry are present, variational 
collapse can occur due to mixing of occupied and virtual orbitals with the same symmetry.
The commonly used self-consistent field (SCF) approach can be combined with 
a maximum overlap method (MOM)\cite{Barca2018, Gilbert2008, Cheng2008}, 
which prevents variational collapse. 
However, SCF convergence can still be problematic when dealing with single determinants
that include unequally occupied degenerate or near-degenerate orbitals. 
This situation is analogous to what happens for ground states with vanishing 
HOMO-LUMO gap\cite{Rabuck1999} and can arise, for example, close to conical 
intersections\cite{Huix-Rotllant2013}. One strategy that is often adopted is 
electronic smearing to obtain convergence on an average occupied configuration\cite{Dickson1996}. 
This, however, comes with the risk of introducing artifacts in the calculated excited-state 
properties, as will be demonstrated below.

There exist alternative nonlinear variational procedures for finding stationary points
of energy functionals based on direct optimization (DO) of the orbitals through unitary 
transformations\cite{Voorhis2002, Hutter1994, Head-Gordon1988, Douady1980}. 
Implemented with gradient-based unconstrained optimization algorithms,
this approach
has been shown to be a more robust strategy for converging ground states with DFT than SCF-based
methodologies\cite{Baarman2011, VandeVondele2003, Voorhis2002}. 
However, the risk of variational collapse impedes straightforward
application of DO methods for locating saddle points of the energy surface.
One way of circumventing this problem is to
convert the saddle-point optimization to a 
minimization of the squared norm of the gradient of the energy with respect to the electronic 
degrees of freedom\cite{Hait2020}. Variational collapse is avoided 
by squared gradient minimization but there is a series of drawbacks that have to be 
considered. First, the computational cost is increased with respect to 
ground-state calculations, because the gradient of the squared norm of the gradient is needed.
Furthermore, this strategy requires more iterations than SCF-MOM 
(when convergence can be reached)\cite{Hait2020}, because squared gradient minimization 
is less well conditioned than energy minimization\cite{Hait2020, Shea2020}.
Lastly, this approach can converge on points where the squared norm of the gradient
has a minimum but is not zero. The initial guess, therefore, needs to be sufficiently good\cite{Hait2020}.

When the above-mentioned practical issues have not represented a problem, excited-state 
calculations using KS functionals have given more accurate results than linear-response TDDFT 
for a number of challenging excited states. 
These include doubly excited states\cite{Hait2020, Barca2018}, core excitations\cite{Hait2020b}, 
Rydberg\cite{Hait2020, Cheng2008} and charge-transfer\cite{Barca2018, Gilbert2008, Liu2017, Briggs2015} 
transitions, absorption spectra\cite{Briggs2013} and structural dynamics\cite{Levi2019, Levi2018} in solution, 
including nonadiabatic dynamics\added[id=Other]{\cite{Malis2020, Pradhan2018, Maurer2011}}. 
However, it has been pointed out\cite{Zhao2019, Gudmundsdottir2013} that many excited states, such as Rydberg, 
charge-transfer and doubly excited states, are affected more by self-interaction error (SIE) than ground states 
at the level of the commonly employed semi-local KS functionals. An unbalanced treatment of self interaction 
can, for example, lead to systematic errors in calculations of excitation energy\cite{Zhao2019}. 
Self-interaction correction (SIC)\cite{Perdew1981} applied to KS functionals corrects the long-range form of the
effective potential, as has been demonstrated, for example, for Rydberg states\cite{Gudmundsdottir2013}
and dipole bound anions\cite{Zhang2016}; thus, it can improve the description of the excited
states\cite{Hemanadhan2014}.
However, it is challenging to perform fully variational calculations with SIC functionals since they 
are explicitly orbital-density dependent and the energy is not invariant to unitary tranformations among 
the equally occupied orbitals. While fully variational implementations of SIC functionals has been developed 
for ground states\cite{Lehtola2016, Borghi2015, Lehtola:2014, Goedecker:1997}, the excited-state calculations 
have so far not been fully variational\cite{Gudmundsdottir2013}.

Here, we present a DO approach with the aim of improving on already existing excited-state DFT methodologies 
in two ways: 
(1) ensuring convergence for different types of excited states, including cases with unequally occupied 
degenerate orbitals, while avoiding variational collapse and without increasing the computational cost
with respect to ground-state DFT calculations; 
(2) allowing the use of non-unitary invariant functionals, such as SIC functionals, in variational excited-state
calculations.
The proposed method uses a quasi-Newton algorithm to directly converge on saddle points of any order with 
the help of a preconditioner built from the eigenvalues of the Hamiltonian matrix at given intervals during 
the optimization, and MOM constraints to prevent variational collapse. 
A preliminary evaluation of the convergence properties 
of the DO-MOM method when using the Limited-memory Broyden-Fletcher-Goldfarb-Shanno (L-BFGS) algorithm 
and a new limited-memory formulation of Powell inverse Hessian update \added[id=Other]{(L-Powell)} is presented in a conference proceeding
\cite{LeviIvanov2020}. L-BFGS is a quasi-Newton method commonly employed for minimization, and it was shown that
the application in the present context
crucially depends on updates of the preconditioner and on the MOM constraints in order to converge on a saddle point. 
\added[id=Other]{L-Powell} was found to be less robust than L-BFGS\cite{LeviIvanov2020}, 
despite its ability to generate indefinite Hessian approximations. It would be advantageous to attain convergence on a
target $n$th-order saddle point without relying on updates of the preconditioner, since it requires costly
diagonalization of the Hamiltonian matrix. In the present work, we extend the limited-memory inverse Hessian
update algorithm presented in reference\cite{LeviIvanov2020} to the symmetric rank-one (SR1) formula. 
SR1 
can develop negative eigenvalues\cite{NocedalWright} and therefore has the 
desired characteristics to minimize the dependency on the preconditioner. 

The convergence properties of the DO-MOM method\cite{LeviIvanov2020}
are tested on \added[id=Rev2]{55 singlet and 34 triplet} excited states of small and medium size molecules, 
including tests of the new limited-memory SR1 \added[id=Other]{(L-SR1)} inverse Hessian update algorithm.
Furthermore, we test the convergence with respect to two challenging charge-transfer states 
of the nitrobenzene molecule, for which SCF-MOM has been reported to fail\cite{Hait2020, Mewes2014}, 
demonstrating that improved robustness and reduced dependency on the preconditioner can be achieved 
with the new \added[id=Other]{L-SR1} method. Finally, we show how the DO-MOM method can converge for systems with 
unequally occupied (near-)degenerate orbitals without tuning modifications, taking 
the conical intersection 
of two excited states of carbon monoxide as a representative example.  
In each case, the performance of DO-MOM is compared to that of a standard SCF-MOM method.

The DO-MOM method can be used for non-unitary invariant functionals such as SIC functionals, as well as the 
unitary invariant KS functionals. We perform fully variational excited-state calculations with SIC on the 
hydrogen atom and dihydrogen molecule and show that the application of SIC in both ground- and excited-state 
calculations leads to significant improvement in the excitation energy.


\section{Theory}\label{theory}
\subsection{Excited-State DFT}
\subsubsection{Kohn-Sham Formulation}
Within KS DFT~\cite{Hohenberg:1964,Kohn:1965}, excited states 
of a spin-polarized system of $N=N_{\uparrow}+N_{\downarrow}$ electrons with
density $n({\bf r})=n_{\uparrow}({\bf r})+n_{\downarrow}({\bf r})$ can be found as saddle points 
of the energy surface defined by the dependence of the ground-state energy on the electronic degrees of freedom\cite{Cheng2008}:
\begin{align}\label{eq:KSDFT1}
E_{\mr{KS}}\left[n_{\uparrow},n_{\downarrow}\right] = T_{\mr s}\left[n_{\uparrow},n_{\downarrow}\right] 
                                                     + V_{\mr{ext}}\left[n\right] 
                                                     + J\left[n\right] 
                                                     + E_{\mr{xc}}\left[n_{\uparrow},n_{\downarrow}\right]
\end{align}
where $T_{\mr s}\left[n_{\uparrow},n_{\downarrow}\right]$ is the kinetic 
energy of the non-interacting $N-$electron system, $V_{\mr{ext}}\left[n\right]$ 
and $J\left[n\right]$ are the energy due to the external potential and the Hartree 
electrostatic energy, respectively:
\begin{align}\label{eq:KSDFT2}
V_{\mr{ext}}\left[n\right] = \int \boldsymbol{\upsilon}_{\mr{ext}}({\bf r})n({\bf r}) d{\bf r}               
\end{align} 
\begin{align}\label{eq:KSDFT3}
J\left[n\right] = \frac{1}{2} \int \int \frac{n({\bf r})n({\bf r}^\prime)}{\mid{\bf r}-{\bf r}^\prime\mid}d{\bf r}d{\bf r}^\prime  
\end{align} 
while $E_{\mr{xc}}\left[n_{\uparrow},n_{\downarrow}\right]$ is the exchange-correlation 
(xc) functional. The KS kinetic energy and the spin densities $n_\sigma({\bf r})$ are given 
in terms of orthonormal KS orbitals $\psi_{n\sigma}({\bf r})$:
\begin{align}\label{eq:KSDFT4}
T_{\mr s}\left[n_{\uparrow},n_{\downarrow}\right] = 
-\frac{1}{2} \sum_{n\sigma} f_{n\sigma} 
\int \psi_{n\sigma}^*({\bf r})\nabla^2 \psi_{n\sigma} ({\bf r}) d{\bf r}       
\end{align}
\begin{align}\label{eq:KSDFT5}
n_\sigma({\bf r}) = \sum_{n} f_{n\sigma}\mid \psi_{n\sigma}({\bf r}) \mid^2      
\end{align}
in which $0 \le f_{n\sigma} \le 1$ is the occupation number for orbital $n$ 
with $\sigma$ spin quantum number ($\uparrow$ or $\downarrow$).

Stationary states of the non-interacting $N-$electron system can be obtained by finding 
extrema of the energy, eq. \ref{eq:KSDFT1}, subject to orbital orthonormality constraints: 
\begin{align}\label{eq:KSDFT6}
\int \psi_{n\sigma}^*({\bf r})\psi_{m\sigma'}({\bf r})d{\bf r} = \delta_{nm}\delta_{\sigma \sigma'}
\end{align}
For a fixed set of $f_{\sigma n}$, the stationarity condition 
leads to a  set of nonlinear coupled equations:
\begin{align}\label{eq:KSDFT7}
f_{n\sigma}{\bf H}_{\mr{KS}}^{\sigma} \psi_{n\sigma} = \sum_{m} \lambda^\sigma_{nm} \psi_{m\sigma}
\end{align}
where the $\lambda^\sigma_{nm}$ are Lagrange multipliers for the constraints, and 
${\bf H}_{\mr{KS}}^{\sigma}$ is the one-particle KS Hamiltonian:
\begin{align}\label{eq:KSDFT8}
{\bf H}_{\mr{KS}}^{\sigma} = -\frac{1}{2} \nabla^2 + \boldsymbol{\upsilon}_{\mr{ext}}({\bf r}) 
                                                 + \int \frac{n({\bf r}^\prime)}{\mid{\bf r}-{\bf r}^\prime\mid}d{\bf r}^\prime  
                                                 + \boldsymbol{\upsilon}_{\mr{xc}}^{\sigma}({\bf r}) 
\end{align}
For a functional with a form given by eq. \ref{eq:KSDFT1}, any unitary 
transformation that mixes equally occupied orbitals among themselves leaves the total energy
unchanged. Therefore, the set of orbitals that makes the 
energy stationary for given set of occupation numbers is not unique.  


\subsubsection{Self-Interaction Correction}
In KS functionals, the Coulomb interaction between the electrons is estimated from the total electron density, and hence  
it includes non-local self interaction.
While the xc functional also includes self interaction of opposite sign, 
a local or semi-local functional form cannot cancel out the self Coulomb interaction and a SIE remains, as can be seen
most clearly for one-electron systems.
Perdew and Zunger\cite{Perdew1981} proposed the following procedure for removing self interaction from a KS functional:
\begin{equation}\label{eq:SIC1}
E_\mathrm{SIC}[n_{\uparrow},n_{\downarrow}] = E_{\mr{KS}}[n_{\uparrow},n_{\downarrow}]
- \sum_{n\sigma} \left( J[n_{n\sigma}] +  E_{xc}[n_{n\sigma}, 0]\right)
\end{equation}
where $n_{n\sigma} = |\psi_{n\sigma}|^{2}$ is an orbital density. 
This represents an orbital-by-orbital estimate of the SIE that is exact for one-electron systems.

For a SIC functional, the stationarity condition leads to a set of nonlinear coupled equations: 
\begin{align}\label{eq:SIC2}
f_{n\sigma}\left({\bf H}^{\sigma}_{\mr{KS}} - {\bf V}_{n\sigma}\right)\phansup \psi_{n\sigma} = \sum_{m} \lambda^\sigma_{nm} \psi_{m\sigma}
\end{align}
where the Hamiltonian contains an orbital-density dependent part:
\begin{align}\label{eq:SIC3}
{\bf V}_{n\sigma} =\int \frac{n_{n\sigma}({\bf r}^\prime)}{\mid{\bf r}-{\bf r}^\prime\mid}d{\bf r}^\prime +  \boldsymbol{\upsilon}_{\mr{xc}}(n_{n\sigma}) 
\end{align}
In contrast to KS functionals, SIC functionals are not invariant 
under unitary transformations of the equally occupied orbitals, and the optimal 
orbitals are uniquely defined as those that extremize the energy of the given SIC functional~\cite{ Lehtola2016, Borghi2015, Lehtola:2014, Klupfel:2012, Messud:2009, Goedecker:1997, Svane:1995}.
This corresponds to maximizing the self-interaction correction,
and involves unitary optimization within the manifold of occupied 
orbitals. 

\subsection{Self-Consistent Field}
For unitary invariant functionals,
eq. \ref{eq:KSDFT7} can be simplified by choosing a unitary transformation 
that diagonalizes $\boldsymbol{\lambda}^\sigma$ while leaving the energy unchanged, leading 
to the generalized KS eigenvalue equations in the canonical form:
\begin{align}\label{eq:KSDFT9}
{\bf H}_{\mr{KS}}^{\sigma}\psi_{n\sigma}({\bf r}) = \epsilon_{n\sigma} \psi_{n\sigma}({\bf r})
\end{align}
For the non-unitary invariant SIC functionals presented in the previous section, 
the Lagrange matrix is not diagonal for the optimal orbitals that extremize the total 
SIC-DFT energy due to the orbital-density dependence\cite{Lehtola2015, Lehtola:2014, Klupfel:2012}.

Solutions to the KS equations are found iteratively, defining the SCF
procedure. 
The ground state corresponds to a minimum of the energy given by the functional and is obtained if 
at each SCF iteration the orbitals are occupied according to the aufbau principle. 
Saddle points on the energy surface are obtained for non-aufbau occupations
and are interpreted as excited states\cite{Lehtola2020, Hait2020, Cheng2008}.
Non-aufbau occupations can be enforced during the 
SCF cycle through the MOM method: at each iteration,
the occupied orbitals are selected as those that overlap most with the occupied orbitals 
of the previous iteration\cite{Gilbert2008} or with a set of fixed reference orbitals\cite{Barca2018, Cheng2008}
(the latter strategy is also known as initial maximum overlap method (IMOM)\cite{Barca2018}). 

\subsection{Direct Optimization}\label{DODFT}
Alternatively, the variational problem can be formulated as an optimization of the
orbitals through application of a unitary transformation to a set of orthonormal
reference orbitals\cite{Douady1980, Head-Gordon1988, Hutter1994, Voorhis2002}:
\begin{align}\label{eq:DM1}
\phi_{p\sigma}({\bf r}) = \sum_{q}U_{pq}^{\sigma} \psi_{q\sigma}({\bf r})
\end{align}
The unitary matrix $\bf{U}$ can be parametrized as the matrix exponential\cite{Hutter1994, Head-Gordon1988}:
\begin{align}\label{eq:DM2}
{\bf U} = e^{\bm \theta}
\end{align}
where $\bm{\theta}$ is required to be anti-Hermitian ($\bm{\theta}=-\bm{\theta}^\dag$) 
in order to preserve the orbital orthonormality. In this way, the energy functional can be directly 
extremized in the linear space formed by anti-Hermitian matrices, which makes it possible to use 
well-established local unconstrained optimization strategies\cite{NocedalWright}. 
The exponential transformation of molecular 
orbitals can be applied to both KS and SIC functionals, since it does not require the
functional to be unitary invariant (unitary optimization for SIC functionals means 
that the elements of $\bm{\theta}$ that mix occupied orbitals are non-zero in contrast 
to KS functionals, as explained in the next section). Moreover, gradient-based direct 
optimization (DO) ensures more rigorous convergence compared to SCF\cite{Lehtola2020, Voorhis2002}. 

For excited states, the unconstrained search can be done with quasi-Newton methods
that are able to locate saddle points. 
Compared to minimization, 
the search for a saddle point is arguably a more challenging 
task, requiring an initial guess that is sufficiently close to the wanted solution and a good approximation 
to the Hessian. Nevertheless, quasi-Newton methods for saddle points have long been employed with
some success in various contexts, most notably transition-state searches on potential 
energy surfaces for atomic rearrangements\cite{Olsen2004, Bofill1994, Culot1992, Baker1986, Simons1983, Cerjan1981}. 
Here, we explore a strategy for DO of saddle points of KS and SIC density functionals 
using quasi-Newton search directions starting from a guess obtained by promoting one or more electrons 
from occupied to unoccupied orbitals of a converged ground-state calculation. 

\section{Implementation}\label{implementation}
We have implemented DO-MOM with KS and SIC functionals in a development branch 
of the Grid-based Projector Augmented Wave (GPAW)\cite{GPAW2, GPAW_LCAO, GPAW1} 
software using localized atomic basis sets to represent the molecular 
orbitals. \added[id=Rev1]{The implementation of the 
exponential transformation for KS functionals is presented in 
Reference\cite{LeviIvanov2020}. A review of this implementation and its 
extension to SIC functionals are given in Appendix \ref{appendix1}.
The MOM is based on a standard implementation using fixed reference orbitals\cite{Barca2018}
as shown in Appendix \ref{appendix3}. In the following, we describe the new L-SR1 
algorithm, including the choice of preconditioner.
}

\subsection{Quasi-Newton Step}\label{quasiNewton}
The computational effort of a quasi-Newton step 
based on updating the Hessian matrix scales as $\mathcal{O}(n^3)$\cite{NocedalWright}, where $n$ is the dimensionality of the 
optimization problem (the present DO implementation based on exponential transformation scales as
$NM$, where $N$ is the number of occupied orbitals and $M$ the number of basis set functions). 
A less computationally demanding approach is to update the inverse Hessian instead of the Hessian, since this 
does not involve any matrix-matrix operation or solution of a linear system of equations. 
The quasi-Newton step with inverse 
Hessian update is:
\begin{align}\label{eq:QuasiNewton1}
{\bf x}^{(k+1)} = {\bf x}^{(k)} - {\bf B}^{(k)}{\bf g}^{(k)}
\end{align}
where ${\bf B}^{(k)}$ is the approximate inverse Hessian at iteration $k$, and ${\bf x}^{(k)}$ and ${\bf g}^{(k)}$ are
the vectors of the $\{\theta_{ij}\}$ independent variables and the analytical gradient, respectively.

When the inverse Hessian is updated, the arithmetic operations scale as
$\mathcal{O}(n^2)$\cite{NocedalWright}, which can become a bottleneck for systems with a moderate 
number of electrons and/or large basis sets.  
To circumvent the costly operations embedded in the explicit update and storage of the Hessian matrix, 
quasi-Newton algorithms can be formulated in a limited-memory version by storing only vectors and 
scalars carrying the information necessary to propagate ${\bf B}$ implicitly. In this case, the operations involved
in one iteration scale linearly as $\mathcal{O}(mn)$, where $m$ is the number of previous steps used to update 
the current step. 

L-BFGS is a commonly used limited-memory version of BFGS, which is generally considered to be the most effective
inverse Hessian update for minimization. The L-BFGS method has been implemented here as
described in reference\cite{NocedalWright}. In calculations of atomic structures, the Powell or SR1
Hessian update formulas, or a combination of the two\cite{Bofill1995, Bofill1994}, are preferred 
for saddle-point searches, because they are able to develop negative eigenvalues contrary 
to the BFGS formula. Therefore, we have formulated and implemented limited-memory variants 
of the Powell and SR1 inverse Hessian updates \added[id=Other]{(L-Powell and L-SR1)} 
by extending the approach based on Powell 
Hessian updates presented by Anglada \emph{et al.}\cite{Anglada1999}. 
\added[id=Rev1]{The L-Powell method is described in reference 
\cite{LeviIvanov2020} and is reviewed in Appendix \ref{appendix2}.}

The inverse Hessian SR1 update formula, written in a compact form, is\cite{SunYuan}:
\begin{align}\label{eq:SR2}
{\bf B}_{\mr{SR1}}^{(k+1)} = {\bf B}^{(k)} + 
                             \frac{{\bf j}^{(k)} {\bf j}^{T(k)}}
                                  {{\bf j}^{T(k)} {\bf y}^{(k)}}
\end{align}
where:
\begin{align}\label{eq:QuasiNewton3}
{\bf j}^{(k)} = {\bf s}^{(k)} - {\bf B}^{(k)}{\bf y}^{(k)}, \quad 
\end{align}
and:
\begin{align}\label{eq:QuasiNewton2}
{\bf s}^{(k)} = {\bf x}^{(k+1)} - {\bf x}^{(k)}, \quad 
{\bf y}^{(k)} = {\bf g}^{(k+1)} - {\bf g}^{(k)}
\end{align}
For any vector ${\bf v}^{(k)}$ and approximation ${\bf B}_0^{(k)}$ 
to the inverse Hessian (${\bf B}_0^{(k)}$ can in principle be allowed to vary at each iteration), 
${\bf B}_{\mr{SR1}}^{(k)}{\bf v}^{(k)}$ can be computed using 
the following recursive formula:
\begin{align}\label{eq:SR3}
{\bf B}_{\mr{SR1}}^{(k)}{\bf v}^{(k)}  = {\bf B}_0^{(k)}{\bf v}^{(k)} + 
                                             \sum_{i=k-m}^{k-1} \frac{{\bf j}^{(i)} {\bf j}^{T(i)}{\bf v}^{(k)}}
                                             {{\bf j}^{T(i)} {\bf y}^{(i)}}
\end{align}
which takes into account the implicit information contained in the $m$ most recent steps. 
Using this result, the \added[id=Other]{L-SR1} algorithm
can be formulated as shown in Algorithm 1.
\begin{algorithm}
\setstretch{1.125}
\SetAlgoNoLine
 Choose  ${\bf x}^{(0)}$, $m$ and $p_{\mr{max}}$\;
 $k \gets 0$\;
 \While{not converged}{
 Choose ${\bf B}_0^{(k)}$\;
 Compute ${\bf p}^{(k)} \gets {\bf B}^{(k)}{\bf g}^{(k)}$ 
                  using eq. \ref{eq:SR3}\;
 \If{$\| {\bf p}^{(k)} \| \ge p_{\mr{max}}$}{
 ${\bf p}^{(k)} \gets \frac{p_{\mr{max}}}{\| {\bf p}^{(k)} \|}{\bf p}^{(k)}$
   }
 ${\bf x}^{(k+1)} \gets {\bf x}^{(k)} - {\bf p}^{(k)}$\;
   \If{$k>m$}{
   discard vector ${\bf j}^{(k-m)}$ 
                 and scalar $r^{(k-m)}$\;
   }
 ${\bf s}^{(k)} \gets {\bf x}^{(k+1)} - {\bf x}^{(k)}$\ and ${\bf y}^{(k)} \gets {\bf g}^{(k+1)} - {\bf g}^{(k)}$\;
 Compute $ {\bf j}^{(k)} \gets {\bf B}^{(k)}{\bf y}^{(k)}$
                  using eq. \ref{eq:SR3}\;
 $ {\bf j}^{(k)} \gets {\bf s}^{(k)} - {\bf j}^{(k)}$\;
 $r^{(k)} \gets {\bf j}^{T(k)} {\bf y}^{(k)}$\;
 Store vector ${\bf j}^{(k)}$ 
 and scalar $r^{(k)}$\;
 $k \gets k + 1$\;
 }
 \caption{Quasi-Newton algorithm with limited-memory SR1 inverse Hessian update. 
 The computational cost of the operations involved scales linearly with 
 $n$ if ${\bf B}_0^{(k)}$ is selected to be diagonal.}
\end{algorithm}
\noindent

Among the quasi-Newton schemes that are commonly used in optimizations of 
saddle points for atomic rearrangements, some update the Hessian through a 
combination of the SR1 and Powell updates. 
Algorithm 1 is readily generalized to use an analogous update formula that 
combines the SR1 and Powell inverse Hessian updates: 
\begin{align}\label{eq:Bofill1}
{\bf B}_{\phi}^{(k+1)} = (1-\phi^k){\bf B}_{\mr{SR1}}^{(k+1)} + \phi^k{\bf B}_{\mr{P}}^{(k+1)}
\end{align}
where ${\bf B}_{\mr{SR1}}^{(k+1)}$ and ${\bf B}_{\mr{P}}^{(k+1)}$ are 
given by eqs. \ref{eq:SR2} and \ref{eq:Powell2}, respectively.
Following Bofill\cite{Bofill1995, Bofill1994}, the factor $\phi^k$ can be taken as:
\begin{align}\label{eq:Bofill2}
\phi^k = 1 - \frac{({\bf y}^{T(k)} {\bf j}^{(k)})^2}
                                  {({\bf y}^{T(k)} {\bf y}^{(k)})({\bf j}^{T(k)} {\bf j}^{(k)})}
\end{align}

In Algorithm 1 we have also introduced a maximum allowed step length, 
$p_{\mr{max}}$. This is because, due to the approximate nature of the initial approximation 
to the Hessian (see next section), initial steps may be too large, causing departure from 
the basin of attraction of the desired saddle point. We have found that $p_{\mr{max}}=0.20$
provides an adequate balance between stability and speed of convergence in most cases.
The SR1 update can become unstable if the denominator in eq. \ref{eq:SR2} is small.
To avoid such instabilities, the following procedure is adopted: 
if $| {{\bf j}^{T(i)} {\bf y}^{(i)}} | < \varepsilon$, where $\varepsilon$ is a small number, 
then ${{\bf j}^{T(i)} {\bf y}^{(i)}}$
is set to $\varepsilon$. When using $\varepsilon=10^{-12}$, we have found that this procedure prevents 
\added[id=Other]{L-SR1} from becoming unstable, without affecting the rate of convergence. 

\subsection{Preconditioner}\label{preconditioner}
The preconditioner for the quasi-Newton step, represented by the matrix ${\bf B}_0^{(k)}$
introduced in the previous section, is chosen as the inverse of the following diagonal
approximation to the Hessian matrix\cite{Head-Gordon1988}:
\begin{align}\label{eq:PREC1}
\frac{\partial^{2} E}{\partial^{2} \theta_{ij}} \approx -2 (\epsilon_{i} - \epsilon_{j} ) (f_{i} - f_{j})
\end{align}
where the $\epsilon_{i}$ are the eigenvalues of the Hamiltonian matrix. 
Eq. \ref{eq:PREC1} is obtained by taking the derivative of a linear expansion of the 
gradient (eq. \ref{eq:MOMDO12}) and neglecting second-order derivatives of the potential. 
Previously, it has been shown that this type of preconditioner can improve the 
convergence of Hartree-Fock (HF) and DFT calculations based on direct minimization of the 
energy\cite{Hait2020,Voorhis2002}, when using the BFGS method.

At the beginning of the optimization, the preconditioner is generated using the 
eigenvalues and occupation numbers of the guess obtained by promoting electrons 
from occupied to virtual orbitals of the ground state. As will be shown for 
the excited states of nitrobenzene, it can happen 
that the number of negative eigenvalues of this initial approximate Hessian is not
consistent with the order of the saddle point corresponding to the target excited
state. To ensure that the preconditioner has the appropriate structure to guide
the convergence towards the target $n$th-order saddle point, the approximate Hessian 
of eq. \ref{eq:PREC1} is recomputed at regular intervals during the optimization 
and ${\bf B}_0^{(k)}$ updated together with the reference orbitals. 
In order to find the occupation numbers of the canonical orbitals, which are needed to 
compute the preconditioner based on eq. \ref{eq:PREC1}, the MOM method is employed
\added[id=Rev1]{(see Appendix \ref{appendix3})}. 
Close to the target solution, the update of the preconditioner is not needed and can be avoided 
using a threshold on the magnitude of the energy gradient, which reduces the computational 
cost by avoiding unnecessary diagonalization of the Hamiltonian matrix.
Finally, we note that the preconditioner derived from eq. \ref{eq:PREC1} 
is not defined for oo terms, since in this case $f_{i}^{(k)} = f_{j}^{(k)}$, and for degenerate 
ov pairs. For these cases, the preconditioner is not used, corresponding to setting the
elements of ${\bf B}_0^{(k)}$ to 1. 


\section{Computational Methods}\label{computational}
All the calculations presented in this work are performed with a development version 
of GPAW where the DO-MOM method for KS and SIC xc functionals has been
implemented. The PAW method\cite{PAW_Blochl:1994} is used to treat the regions near 
the nuclei, core electrons
of each atom are frozen to the result of a reference scalar relativistic calculation of the isolated 
atom, and valence electrons are represented in a basis of linear combination of atomic orbitals.
For all the basis sets considered in this work, the uncontracted functions are removed, 
as the nodal structure of the orbitals around the nuclei is accounted for by the PAW correction.
The simulation cell has a uniform grid with grid spacing of 0.15 \AA{}, while the dimensions 
of the box are chosen in such a way as to avoid effects due to truncation of the numerical representation 
of the basis functions.
For the DO-MOM calculations, a maximum allowed step length, $p_{\mr{max}}$, of 0.20 is utilized, 
while the memory $m$ of L-BFGS, \added[id=Other]{L-Powell and L-SR1} is chosen as equal to 20. 
At every 20th step the preconditioner based on eq. \ref{eq:PREC1} 
is updated unless the root mean square of the gradient is less than 10$^{-3}$ eV. 
\added[id=Rev2]{The SCF-MOM method is based on direct diagonalization of the Hamiltonian 
matrix in the basis of atomic orbitals and updating the electron density 
using a direct inversion in the iterative subspace (DIIS) procedure 
(Pulay mixing of the density\cite{GPAW2, Pulay1980}).
We use GPAW default parameters for the Pulay mixing of the density: 
the number of old densities used in the mixing is 3,
the coefficient used in the linear mixing of the density with the density residual
vector is 0.15, and no damping of short wavelength density changes is used\cite{GPAW2}.}
Unless otherwise stated, convergence (both in SCF or DO calculations) is considered 
achieved if the integrated value of the square of the residuals of 
eq. \ref{eq:KSDFT7} (for KS functionals) or \ref{eq:SIC2} (for SIC functionals)
is less than $4.0 \cdot 10^{-8}$ eV$^2$ per electron. 
All calculations are performed within the spin-unrestricted formalism. Since each state is described by a single determinant, open-shell singlets are not pure-spin states.
Both the KS and SIC calculations use the generalized gradient approximation (GGA) functional PBE\cite{PBE}. 

\subsection{Convergence Tests}
The robustness and rate of convergence of the DO-MOM method is assessed by performing 
single-point calculations of the excited states of small and medium size 
molecules. \added[id=Rev2]{The tests include 52 singlet and 34 triplet excited states 
of 18 small compounds from the benchmark set of reference\cite{Loos2018}, and the 
lowest singlet excited states of 3 medium organic compounds (acetone, benzene and naphtalene) 
from reference\cite{Schreiber2008}, for a total of 89 states generated by single electron transition 
from the ground state. Lowest triplet states have been excluded from 
this benchmark set since they correspond to minima on the energy surface and not 
saddle points.}
These states are chosen
because highly accurate reference data is available making reliable assignment of the states possible, 
and due to the diverse character of the electronic transitions. The 
test set includes \added[id=Rev2]{35} valence (V) 
excitations ($\mr{n}\rightarrow\pi^*$, $\sigma\rightarrow\pi^*$ and $\pi\rightarrow\pi^*$ 
transtions), \added[id=Rev2]{53} Rydberg (R), and \added[id=Rev2]{1} 
charge-transfer (CT) states (the lowest singlet
excited state of hydrogen chloride). The geometries are taken from reference\cite{Loos2018} 
and reference\cite{Schreiber2008}. For the DO-MOM calculations three different 
inverse Hessian update schemes are compared: L-BFGS, \added[id=Other]{L-Powell and L-SR1} (the latter two according to the
limited-memory algorithm presented in section \ref{quasiNewton}). We further compare 
DO-MOM to a standard SCF-MOM method based on direct diagonalization of the Hamiltonian
matrix, as implemented in GPAW\cite{GPAW_LCAO}. For each molecule,  the ground state is first 
converged using SCF. Then, the initial guess for an excited state is generated by a one-electron
excitation involving the occupied and unoccupied orbitals that define the character 
and symmetry of the excited state. 
Convergence is obtained when 
the square of the residuals is less than $10^{-10} \mr{eV}^2$. The maximum number of iterations 
for a calculation is 300. The \emph{aug}-cc-pVDZ basis set\cite{woon1993a, kendall1992a, dunning1989a}
is used.

The calculations of nitrobenzene test both SCF-MOM and DO-MOM with L-BFGS and \added[id=Other]{L-SR1} 
with respect to convergence to the \added[id=Other]{singlet} \added[id=Other]{$^1$}A$_1$(n$_\pi \rightarrow \pi^{\prime*})$ and 
\added[id=Other]{$^1$}A$_1$($\pi^\prime \rightarrow \pi^*$) excited states. Using the ground-state orbitals, the initial guess for the \added[id=Other]{$^1$}A$_1$(n$_\pi \rightarrow \pi^{\prime*})$ state
is generated by promoting an electron from the highest energy $\pi$ lone pair (n$_\pi$) to the second lowest $\pi^*$ 
orbital ($\pi^{\prime*}$), while for the \added[id=Other]{$^1$}A$_1$($\pi^\prime \rightarrow \pi^*$) state excitation is from the second highest occupied 
$\pi$ orbital ($\pi^\prime$) to the lowest unoccupied $\pi^*$ orbital.
The calculations are perfomed at the $C_{2v}$ geometry used in references\cite{Hait2020, Mewes2014}. 
The basis set is def2-TZVP\cite{ahlrics1}, as in the calculations presented in references\cite{Hait2020, Mewes2014}.

To further assess the robustness of the DO-MOM method in cases of 
orbital degeneracy, the potential energy curves (PECs) of the lowest
$^1\Pi(\sigma \rightarrow \pi^*)$ and 
$^1\Delta(\pi \rightarrow \pi^*)$ excited states of carbon 
monoxide are calculated around the conical intersection. The DO-MOM PECs and analytical atomic forces 
are compared with PECs and forces obtained using an 
SCF-MOM method 
where convergence is attained through Gaussian smearing of both the hole and the excited
electron\cite{Levi2018}. Let $N$ denote the number of valence electrons 
described explicitly and $M$ the total number of orbitals included in the calculation.
At each SCF step, the hole $i$ and the excited orbital $a$ are determined through 
the maximum overlap criterion and the occupation numbers of the $n$ lowest 
$N$ orbitals and those of the $m$ orbitals from $N+1$ to $M$ are modified 
according to: 
\begin{align}\label{eq:Smearing1}
f_n(\epsilon_{n})=1-s_{i}(\epsilon_{n}) \\
f_m(\epsilon_{m})=s_{a}(\epsilon_{m})
\end{align}
where $s_{i}(\epsilon_{n})$ and $s_{a}(\epsilon_{m})$ are Gaussian functions 
of the KS eigenvalues:
\begin{align}\label{eq:Smearing2}
&s_{i}(\epsilon_{n})=\dfrac{1}{A_{i}}\exp\left[-\frac{(\epsilon_{n}-\epsilon_{i})^2}{2\sigma^2}\right],
&s_{a}(\epsilon_{m})=\dfrac{1}{A_{a}}\exp\left[-\frac{(\epsilon_{m}-\epsilon_{a})^2}{2\sigma^2}\right]
\end{align}
with the normalization factors being such that the total number 
of electrons is conserved. The width $\sigma$ is chosen as 0.01 eV at
the beginning of the SCF and then it is increased by 0.02 eV every 40th iterations, 
until convergence is reached. 
A similar electronic smearing technique has been 
used before to stabilize the SCF convergence in DFT calculations of PECs\cite{Maurer2011} 
and Born-Oppenheimer molecular dynamics simulations with DFT atomic forces\cite{Levi2019,Abedi2019,Levi2018}.
For all the calculated points of both DO-MOM and SCF-MOM 
PECs, the guess orbitals are from a ground-state calculation
at the reference geometry\cite{Loos2018}, where the interatomic distance 
is 1.134 \AA{}. 
The DO-MOM and SCF-MOM calculations of the PECs of carbon monoxide 
use a dzp basis\cite{GPAW_LCAO} (default in GPAW).

\subsection{Self-Interaction Corrected Calculations}
DO-MOM calculations were carried out of the ground and first three lowest excited states of 
the hydrogen atom and of the ground and $^1\Sigma_g^+(1\sigma_g^2 \rightarrow 1\sigma_u^2)$ doubly 
excited state of the dihydrogen molecule using both PBE and SIC-PBE.
The basis sets are \emph{daug}-cc-pV6Z excluding $g$- and $h$-type functions 
for hydrogen, and \emph{aug}-mcc-pVQZ excluding $f$-type functions for dihydrogen, 
which leads to an excitation energy converged to within $\sim$0.01 eV 
(see Figure S1 in the Supporting Information)
The interatomic distance in dihydrogen is set to 1.4 \AA{} as in reference\cite{Barca2018}.


\section{Results}\label{results}

\subsection{Convergence Tests}
\subsubsection{Benchmarks on Small and Medium Size Molecules}
\begin{figure}[h!]
    \centering
    \includegraphics[width=1\textwidth]{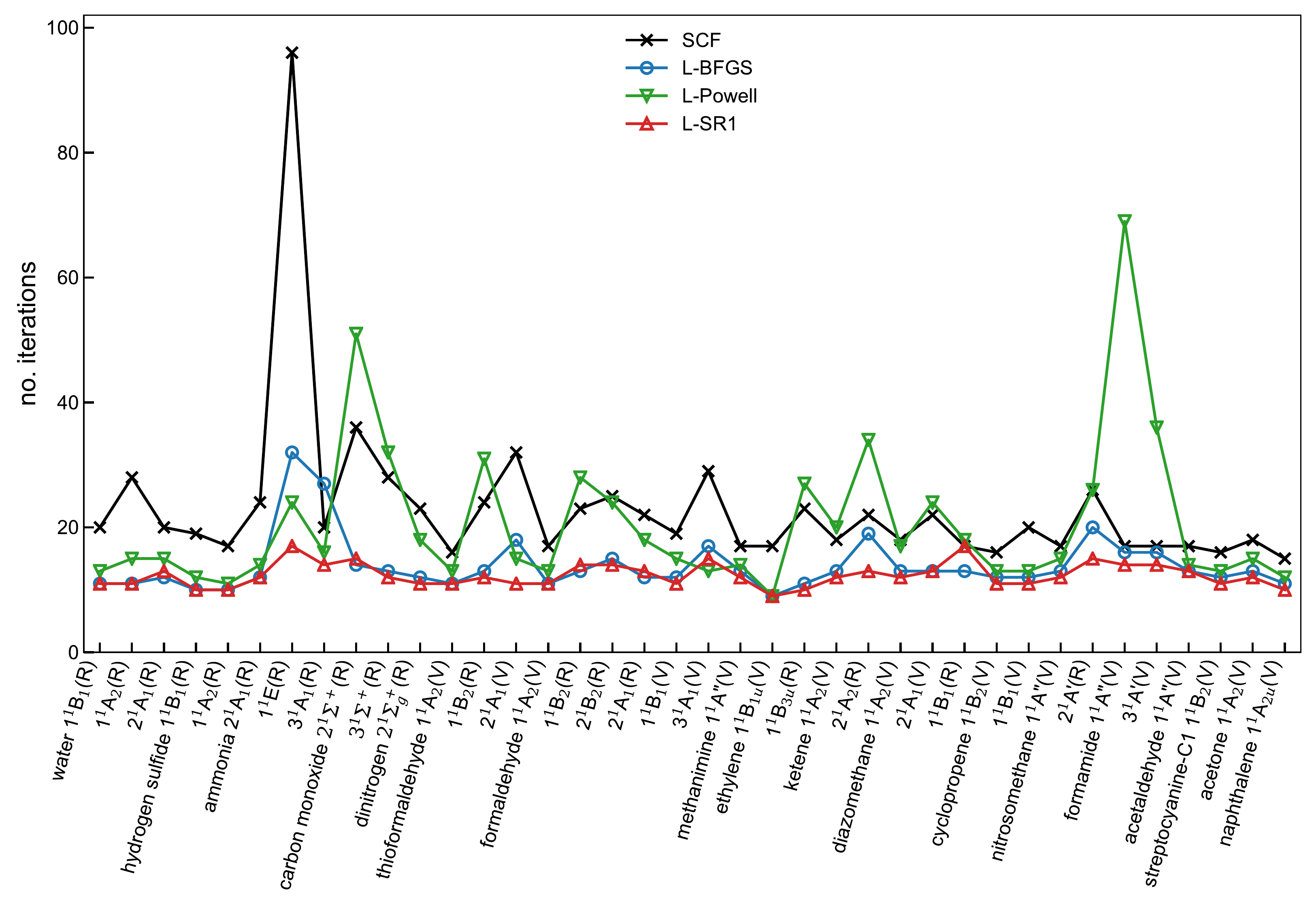}
    \caption{Number of iterations needed to reach convergence of the 
    singlet excited states.
    } 
    \label{fig:Fig1}
\end{figure}

The results of the convergence tests on 55 singlet and \added[id=Rev2]{34 triplet} excited states
of small and medium size molecules are reported in Tables \ref{tbl:Tbl1} and \ref{tbl:Tbl2}.
\begin{table}[h!]
  \small
  \caption{\small Convergence properties of the SCF-MOM and DO-MOM methods
  for 52 singlet excited states of molecules from the benchmark set in reference\cite{Loos2018} plus 
  the lowest excited states of acetone, benzene and naphtalene. For the DO-MOM 
  methods, one iteration corresponds to one energy and gradient evaluation, while for 
  SCF-MOM it represents one energy evaluation and finding the solution for the eigendecomposition of the Hamiltonian matrix.}
  \label{tbl:Tbl1}
  \arrayrulewidth=1.0pt
  \renewcommand{\arraystretch}{1.5}
  \begin{tabular}{p{3.5cm}M{2.0cm}M{2.0cm}M{2.0cm}M{2.0cm}}
    \rowcolor{gray!20}
                 &  SCF-MOM & \multicolumn{3}{c}{DO-MOM} \\
    \hhline{*2{>{\arrayrulecolor{gray!20}}-}*3{>{\arrayrulecolor{black}}-}}
    \rowcolor{gray!20}
                 &                     & L-BFGS     &  L-Powell  & L-SR1   \\
    Convergence failures                & 17     & 2       &  10      &	 0  \\
    \makecell[l]{Avg no. iterations}     & 22.9  & 13.9	& 20.5 & 12.3  \\
    \makecell[l]{Max no. iterations}    & 96     & 32    &  69    &	17	  \\
    \makecell[l]{Min no. iterations}     & 15     & 9       &  9      &	 9  \\
    \makecell[l]{Local saddle points} & 0       & 1       &  1      & 1 \\
  \end{tabular} \\
\end{table}
\begin{table}[h!]
  \small
  \caption{\small Convergence properties of the SCF-MOM and DO-MOM methods
  for 34 triplet states of molecules from the benchmark set in reference\cite{Loos2018}. 
  The calculations corresponding to one iteration 
  are the same as in Table \ref{tbl:Tbl1}.}
  \label{tbl:Tbl2}
  \arrayrulewidth=1.0pt
  \renewcommand{\arraystretch}{1.5}
  \begin{tabular}{p{3.5cm}M{2.0cm}M{2.0cm}M{2.0cm}M{2.0cm}}
    \rowcolor{gray!20}
                 &  SCF-MOM & \multicolumn{3}{c}{DO-MOM} \\
    \hhline{*2{>{\arrayrulecolor{gray!20}}-}*3{>{\arrayrulecolor{black}}-}}
    \rowcolor{gray!20}
                 &                     & L-BFGS     & L-Powell   & L-SR1   \\
    Convergence failures                 & 10     & 0       &  7    &	 0   \\
    \makecell[l]{Avg no. iterations}     & 26.6  & 15.1	& 	23.0 & 12.4  \\
    \makecell[l]{Max no. iterations}    & 121    & 35    &  44  &	 16  \\
    \makecell[l]{Min no. iterations}     & 15     & 9       &  11    &	 10  \\
    \makecell[l]{Local saddle points} & 0       & 2       & 1      & 4 \\
  \end{tabular} \\
\end{table}
The average, maximum and minimum number of iterations are 
reported after excluding the cases that do not converge for any of the methods.
Figure \ref{fig:Fig1} shows the number of iterations needed to 
converge the singlet states.

\begin{figure}[h!]
    \centering
    \includegraphics[width=0.5\columnwidth]{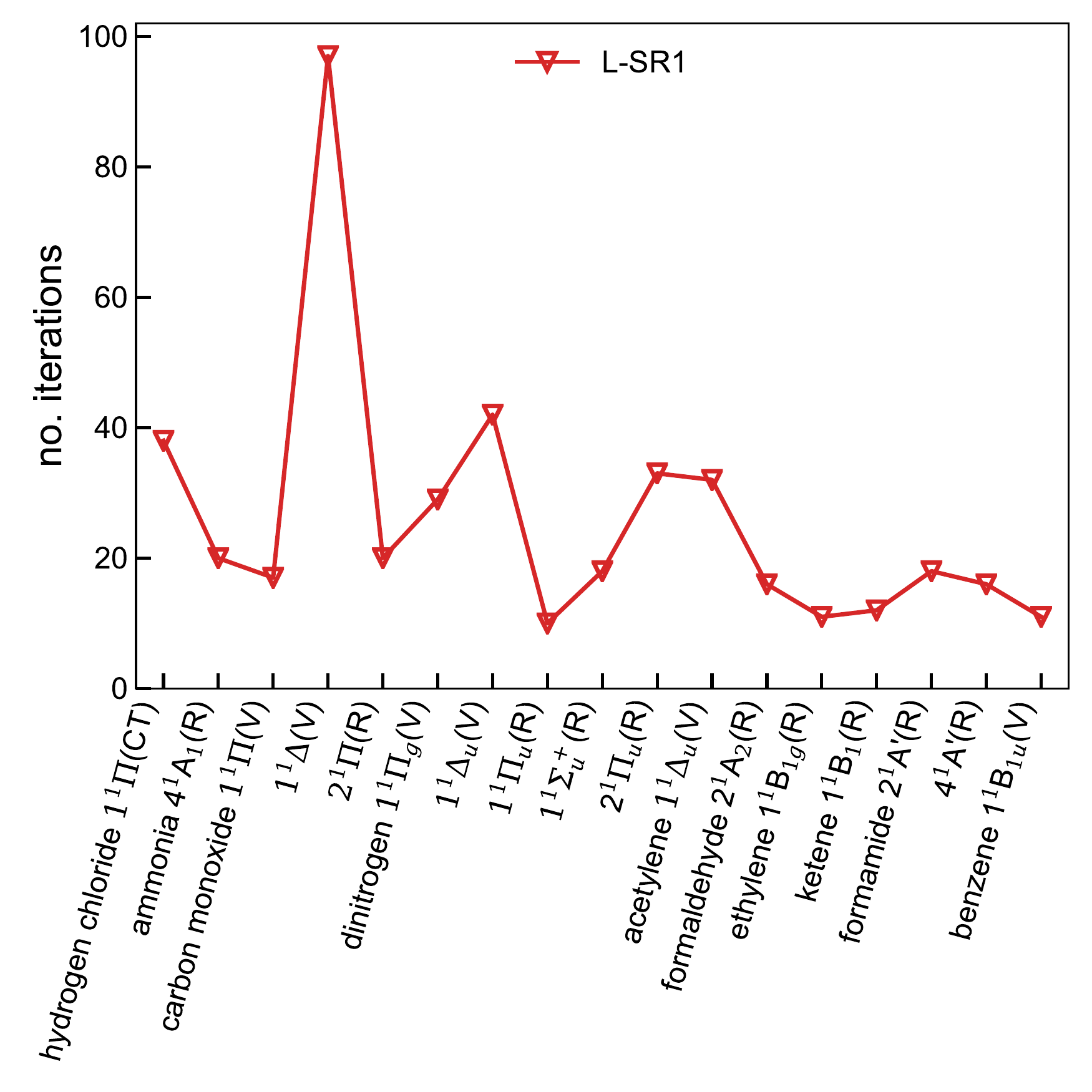}
    \caption{Total number of iterations when DO-MOM with \added[id=Other]{L-SR1} 
    is used to converge the singlet excited states for which SCF-MOM fails.} 
    \label{fig:Fig2}
\end{figure}

SCF-MOM fails to converge within the maximum number of iterations threshold  
in about 30\% of the cases.
All the quasi-Newton algorithms employed within the DO-MOM 
framework are more robust and show a faster rate of convergence than 
SCF-MOM. The best performance is obtained with \added[id=Other]{L-SR1}, 
for which all calculations converge, and 
convergence takes on average about 
\added[id=Rev2]{11 and 14 iterations less than 
SCF-MOM for singlet and triplet states, respectively.}
L-BFGS also performs well, being able to converge in all cases except  
two (the $^1\Delta(\pi \rightarrow \pi^*)$ states of carbon monoxide and dinitrogen). 
\deleted[id=Rev1]{We note that in some cases a gradient along a direction of negative
curvature can be zero because the orbitals forming the corresponding ov 
pair belong to different irreducible representations of the molecule 
symmetry point group. 
This happens in particular for first-order saddle 
points, in which only one ov pair is associated with a direction of negative
curvature (an exception is represented by the lowest excited state of
ammonia, which has A$_1$ symmetry). 
This occurrence may in part 
explain the good performance of L-BFGS, which is in general 
more suited for minimization.}
\added[id=Rev1]{The limited-memory 
Powell inverse Hessian update in DO-MOM is considerably less efficient 
than L-SR1 and L-BFGS. L-Powell can have a slow rate  
of convergence close to a stationary point, which in many cases 
precludes convergence within the maximum number of allowed iterations.}
\added[id=Rev1]{We have tested different combinations of 
limited-memory inverse Hessian updates 
by considering some of the cases that are most difficult to converge 
(see Figure \ref{fig:Fig2}). 
The combination of L-SR1 and L-Powell according to
the Bofill approach (see eqs. \ref{eq:Bofill1} and \ref{eq:Bofill2}) 
is found to have similar performance as L-Powell.} This is 
consistent with the fact that in the Bofill approach \added[id=Rev1]{$\phi^k$}, 
representing the weight of the Powell update, tends to 1 when the optimization approaches 
a stationary point. \added[id=Rev1]{We have also tested a combination of L-SR1 and L-BFGS 
updates using the Bofill factor\cite{Hratchian2005}. This approach
does not lead to better convergence compared 
to L-SR1.}

Figure \ref{fig:Fig2} shows the number of iterations needed by
DO-MOM with SR1 update to converge the singlet excited states for which 
the other methods fail. 
On average these difficult cases require more iterations 
than the cases presented in Figure \ref{fig:Fig1}. Among the states that are difficult to 
converge are those where excitation occurs from or to a degenerate 
pair of $\pi$ orbitals, such as the $\Pi$ 
states of hydrogen chloride, carbon monoxide and dinitrogen, while others are high-lying Rydberg 
states, most of which involve near-degenerate p-type Rydberg orbitals.
\added[id=Rev2]{A particularly challenging situation arises when both the donor and 
the acceptor orbitals involved in the excitation belong to degenerate 
pairs, as for the $\Delta(\pi \rightarrow \pi^*)$ states 
of carbon monoxide and dinitrogen. In this case, all methods except L-SR1 fail to 
converge. The other DO-MOM methods exhibit oscillations between different critical 
points, failing to converge to the desired solution. The failure of SCF when
degenerate orbitals are unequally occupied is discussed in detail 
below.
}
\added[id=Rev1]{In about 30\% of the calculations that do not converge 
with SCF-MOM, 
occupied orbitals can mix with lower-lying empty orbitals 
of the same symmetry.
On the other hand, the L-SR1 method is able to 
converge all these cases.
The properties of DO-MOM when orbitals involved in the excitation are 
allowed to mix are analyzed in more detail in the following 
section, where calculations of two totally symmetric excited states of the 
nitrobenzene molecule are presented.
}

The SCF convergence problems for states where degenerate orbitals are unequally
occupied arise because
the electron density represented by a single determinant 
of KS orbitals obtained at each step is not well defined and this can lead to 
large changes in the orbitals involved in the excitation 
from one step to another, and oscillations between different critical points\cite{Voorhis2002}. 
Several options can be used
to help convergence of the SCF: different mixing of the density, 
\added[id=Rev2]{alternative DIIS extrapolation techniques\cite{Garza2012},}
or tuning modifications such as damping, level shifting of the iterations, or electronic smearing
\cite{Rabuck1999}. 
However, DO is able to follow the same solution more consistently
without such 
modifications. 
It can obtain convergence in these difficult 
cases if the chosen quasi-Newton method guarantees sufficiently accurate 
Hessian updates for the given form of the preconditioner, as shown here.
The robustness of the DO approach in calculations involving orbital
degeneracies has been previously recognized 
for ground states of systems with vanishing HOMO-LUMO gap\cite{Voorhis2002}.

For some excited states, a DO method can converge on a solution with 
higher energy than the solution obtained by SCF or another DO method
for the same excited state. For example, the solution obtained for the 
1$^1$E(n $\rightarrow$ 3p) state of ammonia with L-BFGS or \added[id=Other]{L-SR1} DO-MOM
lies $\sim$0.03 eV higher in energy with respect to the SCF-MOM solution. 
The occurrence of higher-energy solutions, which we refer to as ``local saddle points'', is
indicated in Tables \ref{tbl:Tbl1} and \ref{tbl:Tbl2}. We stress that the 
multiple solutions that are obtained for a particular case are all saddle points 
of the same order and correspond to the same excited state. 
\added[id=Rev2]{Multiple solutions corresponding to the same excited state 
are found to differ in the orientation of the 
highest occupied molecular orbitals (see Figures S2 to S6 in the Supporting Information).}
Similar to what is observed here for saddle points, the geometric direct minimization 
method of reference\cite{Voorhis2002} exhibits a tendency to converge on
local minima of energy functionals compared to SCF minimizers. 
Defining the ``optimal'' approximation to an excited state among multiple
variational solutions might not be trivial.
Indeed, variational solutions of a nonlinear optimization are in general 
not orthogonal to one another, and hence higher solutions are not 
necessarily upper bounds to the exact excited states, but only
upper bounds to the ground state\cite{MolecularElectronicStructureTheory}.
Besides, for many practical applications, such as calculations of PECs or 
molecular dynamics, one is usually only interested in consistently 
converging on the same stationary point.
For these cases, DO-MOM can be 
used without modifications. For cases in which the lowest energy 
saddle point of a given excited state is desired, a possible strategy 
could be to combine the DO approach with techniques for guiding the 
convergence towards a global solution, such as the one presented in 
reference \cite{Vaucher2017}.

Finally, we note that due to the small size of the molecules considered here, 
the computational effort of SCF-MOM and DO-MOM is comparable, as
indicated by similar values of the elapsed time per iteration. For larger systems, 
care must be taken that the memory of the quasi-Newton algorithm used 
within DO-MOM, which here is chosen as $m=20$, does not degrade the 
computational performance of the method. From test calculations, where we 
compare the convergence of L-BFGS and \added[id=Other]{L-SR1} with different levels of memory, 
we find that \added[id=Other]{L-SR1} tends to become less robust with lower memory faster 
than L-BFGS. Therefore, for large systems, L-BFGS might represent the 
best compromise between speed of convergence and computational effort
among the various limited-memory inverse Hessian update schemes. 

\subsubsection{Nitrobenzene}
\begin{figure}[h!]
    \centering
    \includegraphics[width=0.75\columnwidth]{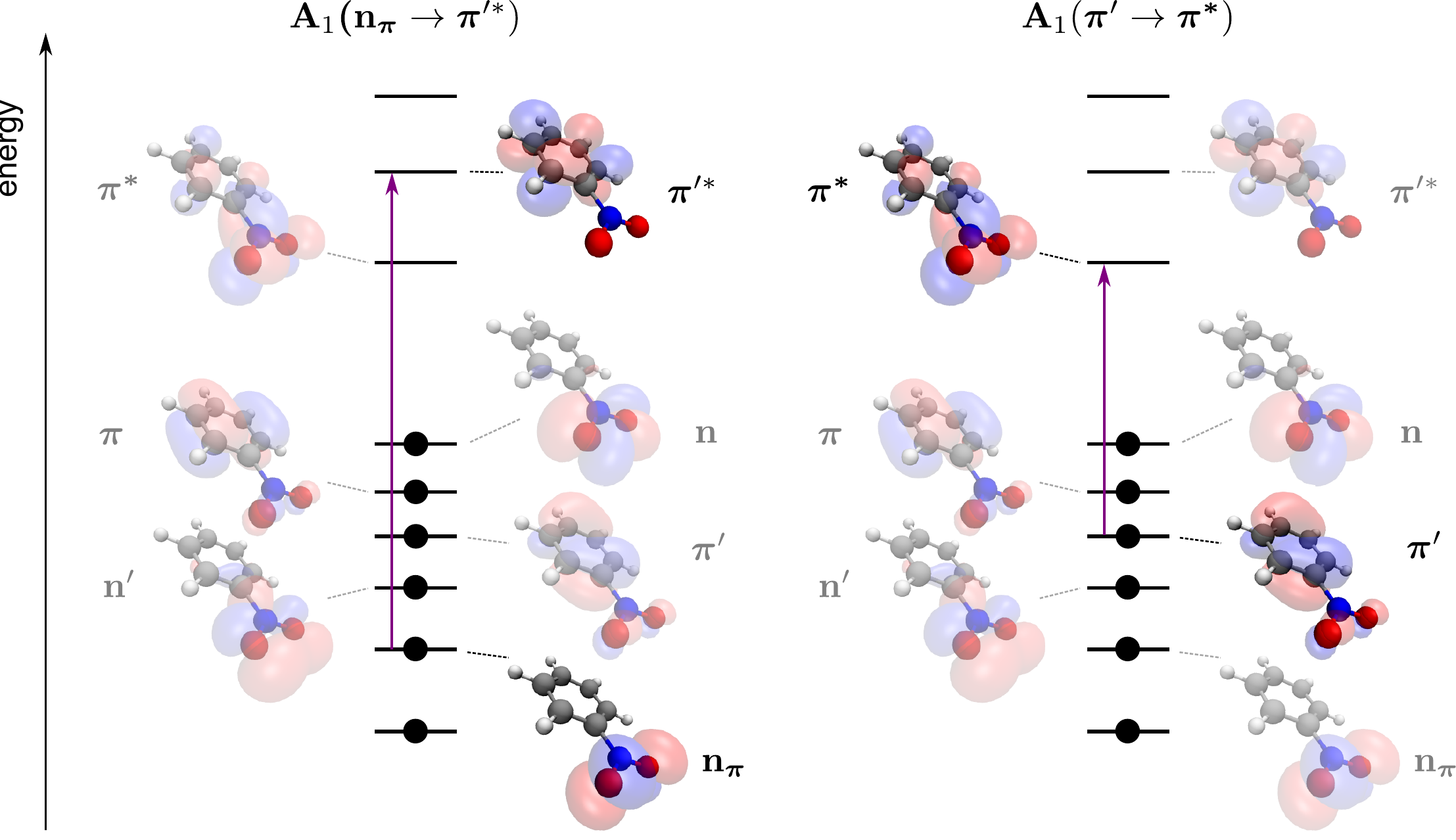}
    \caption{Ground-state frontier molecular orbitals of nitrobenzene and depiction of the electronic
    transitions involved in the \added[id=Other]{$^1$}A$_1$(n$_\pi \rightarrow \pi^{\prime*})$ (Left) and 
    \added[id=Other]{$^1$}A$_1$($\pi^\prime \rightarrow \pi^*$) (Right) excited states. The labels of the orbitals are 
    according to the notation from reference\cite{Mewes2014}. The orbital surfaces are drawn at an 
    isovalue of 0.1 \AA{}$^{-3/2}$.} 
    \label{fig:Fig4}
\end{figure}

Figure \ref{fig:Fig4} illustrates the frontier molecular orbitals involved in the electronic transitions 
that lead to the \added[id=Other]{$^1$}A$_1$(n$_\pi \rightarrow \pi^{\prime*})$ and \added[id=Other]{$^1$}A$_1$($\pi^\prime \rightarrow \pi^*$) 
excited states of nitrobenzene. 
Both states have charge-transfer character: in the case of the \added[id=Other]{$^1$}A$_1$(n$_\pi \rightarrow \pi^{\prime*})$ 
state, one electron moves from the nitro group to the benzene ring, while in the case of the 
\added[id=Other]{$^1$}A$_1$($\pi^\prime \rightarrow \pi^*$) state, the direction of the charge transfer is \added[id=Other]{reversed}. 
Figure \ref{fig:Fig4} also schematically illustrates 
that the highest occupied orbitals, including the orbital from which excitation occurs, are all 
closely spaced in energy, covering a range of around 1 eV, despite being localized on different 
regions of the molecule. Charge transfer from such a subset of closely spaced orbitals is expected 
to be accompanied by a change of the energy ordering of the occupied orbitals.

Hait \emph{et al.}\cite{Hait2020} and Mewes \emph{et al.}\cite{Mewes2014} have shown 
that SCF-MOM-based techniques fail to converge to the \added[id=Other]{$^1$}A$_1$(n$_\pi \rightarrow \pi^{\prime*})$ 
and \added[id=Other]{$^1$}A$_1$($\pi^\prime \rightarrow \pi^*$) states, respectively.
When the overlaps used to find the occupation numbers with the MOM at one step 
are computed with respect to the orbitals from the previous step, collapse to the 
ground state occurs; while if the overlaps are computed with respect to the initial 
set of orbitals, the iterative procedure does not converge. 
In accord with this, our SCF-MOM calculations exhibit large and rapid oscillations 
without convergence in 300 iterations. This failure is likely caused by the presence of 
orbitals energetically close to the n$_\pi $ and $\pi^\prime$ orbitals from which excitation 
occurs, and to rearrangements in the order of the orbital energy levels. 
DO-MOM, however, is able to converge both of these challenging cases. 

\begin{figure}[h!]
    \centering
    \includegraphics[width=0.5\columnwidth]{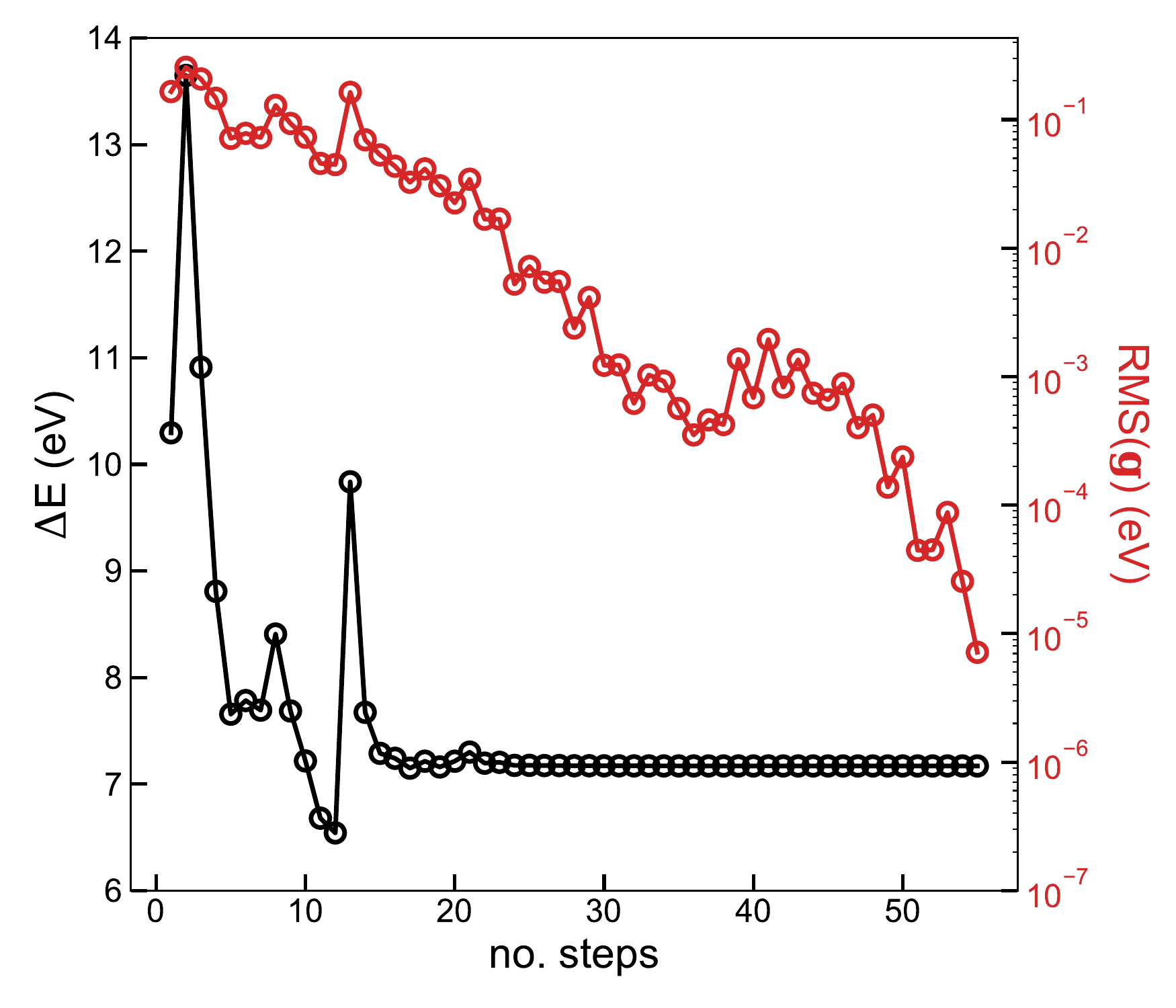}
    \caption{Convergence of excitation energy and root mean square of the gradient in a DO-MOM 
    calculation of the \added[id=Other]{$^1$}A$_1$(n$_\pi \rightarrow \pi^{\prime*})$ excited state of nitrobenzene 
    using L-BFGS.} 
    \label{fig:Fig5}
\end{figure}

\begin{figure}[h!]
    \centering
    \includegraphics[width=1\columnwidth]{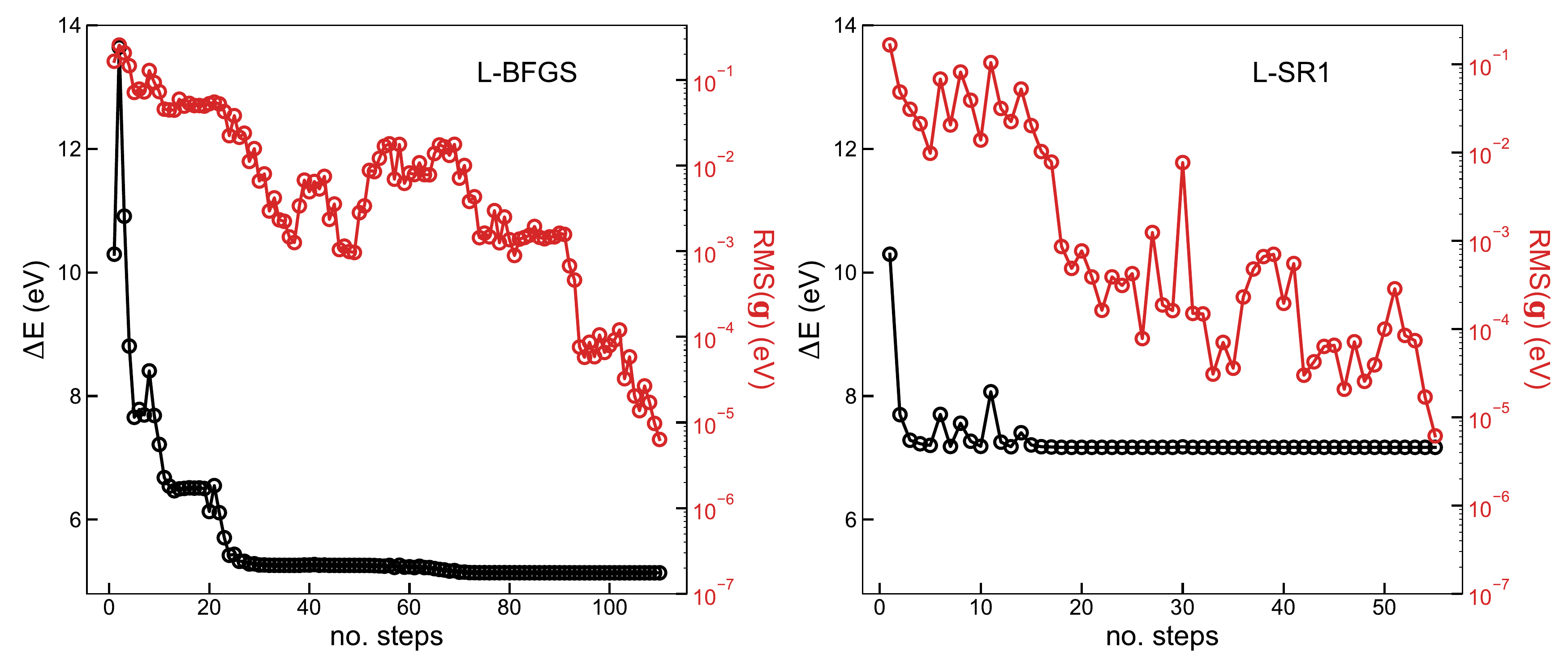}
    \caption{Convergence of excitation energy and root mean square of the gradient in DO
    calculations using L-BFGS (Left) and \added[id=Other]{L-SR1} (Right) without the MOM 
    and with a preconditioner fixed at the guess for the \added[id=Other]{$^1$}A$_1$(n$_\pi \rightarrow \pi^{\prime*})$ state of nitrobenzene.} 
    \label{fig:Fig6}
\end{figure}

Figure \ref{fig:Fig5} shows the convergence of energy and gradient in a DO-MOM 
calculation of the \added[id=Other]{$^1$}A$_1$(n$_\pi \rightarrow \pi^{\prime*})$ state using the L-BFGS method, where 
the preconditioner is updated after the MOM determines a change in the occupation numbers. 
After 13 steps of the optimization, a change of the character of the occupied orbitals 
is detected and, as a result, the MOM induces a 
change
in the occupation numbers, which 
restores the character of the initial guess. Application of the MOM constraints is accompanied 
by a jump in the energy as can be observed from Figure \ref{fig:Fig5}.
After that, the energy is converged to 10$^{-6}$ eV in $\sim$50 steps. 
While the approximate Hessian at the initial guess has six negative eigenvalues, the converged 
solution is a ninth-order saddle point. This is a consequence of a significant rearrangement in 
the ordering of the orbitals induced by the charge transfer, which stabilizes the orbitals 
localized on the nitro group, including the hole, and destabilizes the orbitals localized 
on the benzene ring. When L-BFGS is used, it is essential to apply the MOM constraints and update the 
preconditioner in order to achieve convergence to the target excited state.
This is illustrated in Figure \ref{fig:Fig6}, which shows a DO calculation with L-BFGS starting from the same 
initial guess as in Figure \ref{fig:Fig5} but where the MOM is not applied and the preconditioner 
is not updated. 
In this case,  the hole which has an initial n$_\pi $ character, acquires during the DO the character 
of the $\pi$ orbital depicted in Figure \ref{fig:Fig4} (the n$_\pi $ and $\pi$ orbitals are allowed to mix because 
they both belong to the A$_2$ irreducible representation in the $C_{2v}$ point group symmetry), and the 
calculation eventually collapses to a third-order saddle point. Figure \ref{fig:Fig6} also shows a DO calculation
without the MOM and with a fixed preconditioner when the approximate inverse Hessian is updated using 
\added[id=Other]{L-SR1}. Despite an initial approximate Hessian 
with a lower number of negative eigenvalues compared to the Hessian of the target solution, the DO with \added[id=Other]{L-SR1} is able 
to converge to the ninth-order saddle point corresponding to the \added[id=Other]{$^1$}A$_1$(n$_\pi \rightarrow \pi^{\prime*})$ state.
This can be explained with the ability of \added[id=Other]{L-SR1} to develop negative eigenvalues, while L-BFGS cannot.
The squared gradient minimization method of reference\cite{Hait2020}
is also able to converge to the \added[id=Other]{$^1$}A$_1$(n$_\pi \rightarrow \pi^{\prime*})$ state of nitrobenzene. However, 
due to the need to compute the derivative of the squared norm of the gradient at each step, the 
minimization involves larger computational effort per iteration than the present DO-MOM calculations.

\begin{figure}[h!]
    \centering
    \includegraphics[width=1\columnwidth]{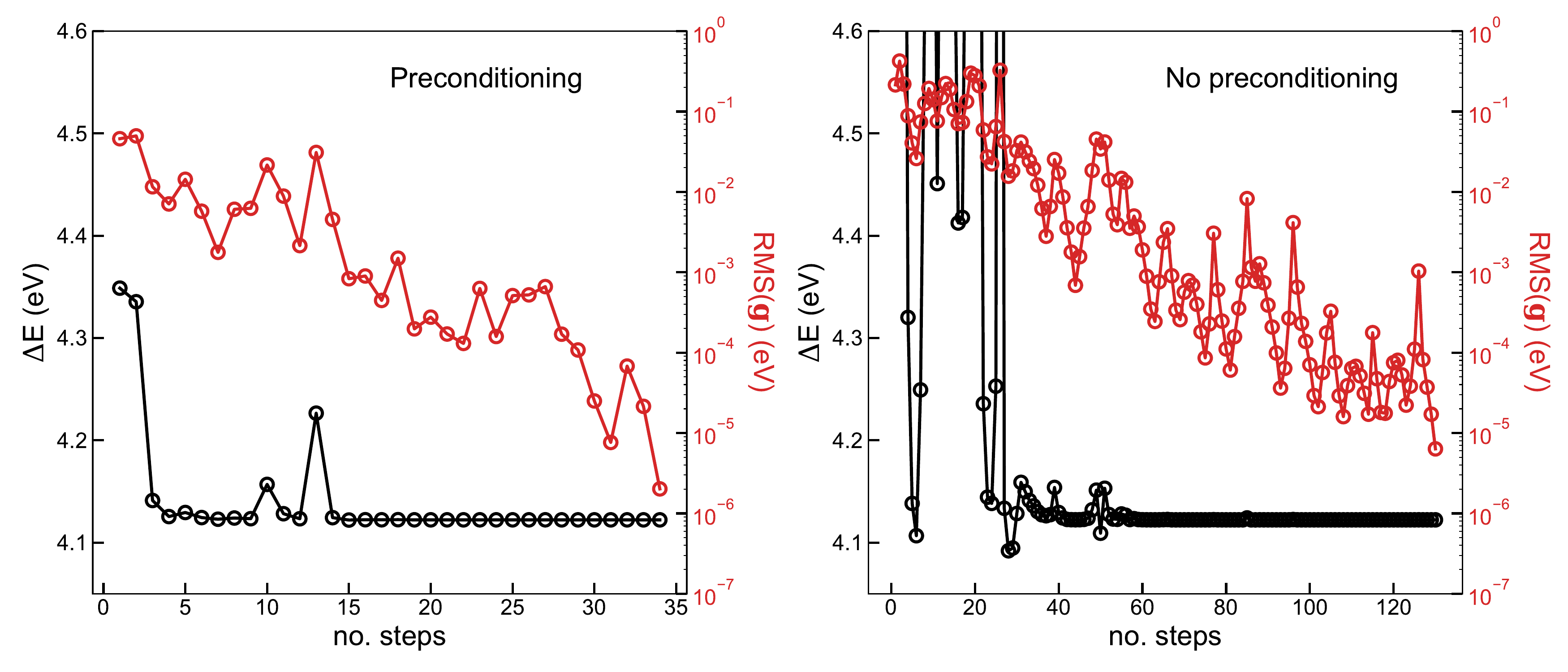}
    \caption{Convergence of excitation energy and root mean square of the gradient in DO-MOM 
    calculations of the \added[id=Other]{$^1$}A$_1$($\pi^\prime \rightarrow \pi^*$) excited state of nitrobenzene 
    using \added[id=Other]{L-SR1} with (Left) and without (Right) preconditioner.} 
    \label{fig:Fig7}
\end{figure}

In the case of the \added[id=Other]{$^1$}A$_1$($\pi^\prime \rightarrow \pi^*$) excited state, the converged 
solution is found to be a fourth-order saddle point, while the approximate Hessian at the initial guess 
generated from the ground-state orbitals (see Figure \ref{fig:Fig4}) has three negative eigenvalues. 
As for the \added[id=Other]{$^1$}A$_1$(n$_\pi \rightarrow \pi^{\prime*})$ state, a DO calculation with 
L-BFGS can converge to the target solution only if the MOM is used and the preconditioner updated during the optimization
\cite{LeviIvanov2020}. Figure \ref{fig:Fig7} shows the convergence of DO-MOM calculations using
\added[id=Other]{L-SR1} with and without preconditioner.
It is found that DO-MOM with \added[id=Other]{L-SR1} is able to converge to the \added[id=Other]{$^1$}A$_1$($\pi^\prime \rightarrow \pi^*$)
state even without the use of a preconditioner, although large oscillations of the energy are observed 
at the beginning of the optimization and almost four times as many steps are required to achieve convergence 
compared to a calculation that uses the preconditioner. These results show that the \added[id=Other]{L-SR1}
method developed in the present work is less sensitive to the quality of the preconditioner and is able to build a 
better approximation to the inverse electronic Hessian when used in optimizations of excited states within DFT compared 
to a standard implementation of the most used L-BFGS quasi-Newton algorithm. 

All calculations presented above use a maximum step length, $p_{\mr{max}}$, of 0.20, 
which is the value found optimal in most of the cases. However, a $p_{\mr{max}}$ of 0.25
leads to smaller oscillations at the beginning of the optimization and faster convergence in the case of the 
DO-MOM calculations with \added[id=Other]{L-SR1} (see Figures S7 and S8 in the Supporting Information). The use of a fixed allowed 
step length is a limitation of the current implementation. To ensure smooth and 
monotonic convergence for a broad range of systems, a trust region scheme could 
be introduced.

\subsubsection{Potential Energy Curves of Carbon Monoxide}
\sloppy The electron configuration of the ground state of carbon monoxide is 
$ (1\sigma)^2(1\sigma^*)^2(2\sigma)^2(2\sigma^*)^2(1\pi)^4(3\sigma)^2(1\pi^*)^0(3\sigma^*)^0 $.
The lowest singlet excited states arise from $\sigma\rightarrow\pi^*$ 
and $\pi\rightarrow\pi^*$ single-electron excitations. Among the states 
with these configurations, the $1^1\Pi$\added[id=Other]{(n$\rightarrow\pi^*$)} 
and $1^1\Delta$\added[id=Other]{($\pi\rightarrow\pi^*$)} can be approximated using a single 
determinant. 

KS DFT has several difficulties describing the
$1^1\Pi$\added[id=Other]{(n$\rightarrow\pi^*$)}  and $1^1\Delta$\added[id=Other]{($\pi\rightarrow\pi^*$)} states and their conical intersection. 
Firstly, the determinant obtained from a single-electron transition between 
orbitals of the same spin has a broken spin symmetry, since the pure
singlet open-shell state is a symmetry-adapted linear combination of two
determinants with the same configuration. Secondly, KS DFT neglects the 
multireference character of the wave functions arising from mixing of 
configurations involving the degenerate pairs of $1\pi$ and $1\pi^*$ orbitals.
Finally, at the conical intersection the $1\pi$ orbitals become degenerate with the 
$3\sigma$ orbital, further increasing the multireference character of the states.
The strong static correlation 
prevents the SCF-MOM method with integer occupation numbers 
from converging. 
Convergence can be achieved 
by smearing the hole and excited electron over the degenerate orbitals.
We emphasize that the aim here is not to assess the accuracy of DFT 
with KS functionals in the description of the excited states, 
for which highly accurate multireference calculations are available 
when the molecules are small, 
but rather to demonstrate the ability of the DO-MOM method to handle 
a challenging case without \emph{ad hoc} modifications to achieve convergence. 

\begin{figure}[h!]
    \centering
    \includegraphics[width=0.65\columnwidth]{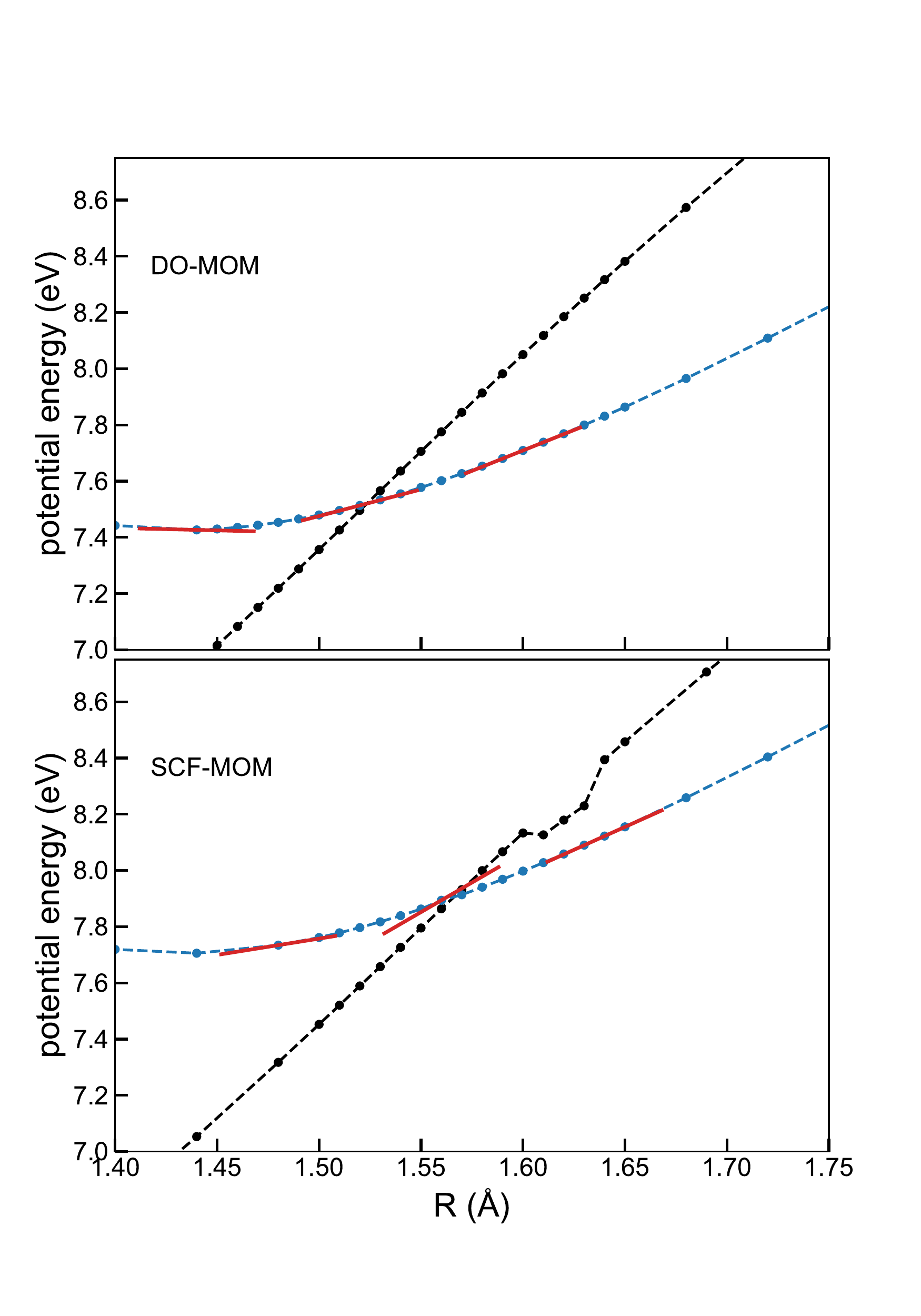}
    \caption{Potential energy curves of the $1^1\Pi$\added[id=Other]{(n$\rightarrow\pi^*$)}  (black) and $1^1\Delta$\added[id=Other]{($\pi\rightarrow\pi^*$)} (blue) excited states of carbon monoxide
    computed with the DO-MOM method (top) and a variant of SCF-MOM using Gaussian smearing 
    of the hole and excited electron (bottom). The red lines represent the analytical atomic
    forces at selected points.} 
    \label{fig:Fig3}
\end{figure}

The PECs of the $1^1\Pi$\added[id=Other]{(n$\rightarrow\pi^*$)} and $1^1\Delta$\added[id=Other]{($\pi\rightarrow\pi^*$)} states of carbon monoxide 
around the conical intersection computed using SCF-MOM with Gaussian smearing and 
DO-MOM are shown in Figure \ref{fig:Fig3} together with the analytical atomic 
forces for selected points on the $1^1\Delta$\added[id=Other]{($\pi\rightarrow\pi^*$)} curves.
For the Gaussian smearing SCF-MOM calculations, far from the conical intersection, the occupation 
numbers of the $1\pi$ orbitals are 1 ($1^1\Pi$\added[id=Other]{(n$\rightarrow\pi^*$)}  state) or 0.5 ($1^1\Delta$\added[id=Other]{($\pi\rightarrow\pi^*$)}), while the 
occupation of $3\sigma$ is either 0 ($1^1\Pi$\added[id=Other]{(n$\rightarrow\pi^*$)} )  or 1 ($1^1\Delta$\added[id=Other]{($\pi\rightarrow\pi^*$)}). Close to 
the conical intersection, the hole can be smeared over both the $1\pi$ and $3\sigma$ orbitals (see 
Tables S3 and S4 in the Supporting Information). When this happens, 
the SCF-MOM PECs display some artefacts. The PEC of $1^1\Pi$\added[id=Other]{(n$\rightarrow\pi^*$)}  shows discontinuities around
the three points for which the smearing is largest. The only point of the 
PEC of $1^1\Delta$\added[id=Other]{($\pi\rightarrow\pi^*$)} for which the hole is smeared over three orbitals coincides approximately 
with the point of crossing of the two curves (R$\sim$1.56 \AA{}). The analytical forces
computed at this point are not consistent with the slope of the $1^1\Delta$\added[id=Other]{($\pi\rightarrow\pi^*$)} curve.

On the other hand, the curves obtained with DO-MOM PECs and integer occupation 
numbers do not exhibit discontinuities and the computed atomic forces are  
consistent with the slopes of the curves. We further note that Gaussian smearing SCF-MOM converges on higher-energy solutions, with the $1^1\Pi$\added[id=Other]{(n$\rightarrow\pi^*$)}  
and $1^1\Delta$\added[id=Other]{($\pi\rightarrow\pi^*$)} PECs computed with DO-MOM lying respectively $\sim$0.1 
and $\sim$0.3 eV lower in energy with respect to the SCF-MOM PECs. 
This also affects the relative position of the two conical intersections.

Electronic smearing is often employed together with the SCF method to converge 
excited-state DFT calculations, especially in molecular dynamics simulations
\cite{Abedi2019, Pradhan2018, Levi2018, Dohn2016, dohn2014}. 
The results presented here show that one needs to carefully check 
whether artefacts are introduced due to the smearing.
The DO-MOM method can converge energy and forces 
even in cases of degeneracies without the need of smearing.

\subsection{Calculations with Self-Interaction Correction}
\subsubsection{Hydrogen Atom}
Table \ref{tbl:Tbl3} reports the energy and eigenvalue of the occupied orbital 
for the ground state and each of the three lowest excited states of the hydrogen 
atom computed with PBE and SIC-PBE using DO-MOM, as compared to the experimental 
values of the ionization energy.  
\begin{table}[h!]
  \centering
  \small
  \caption{\small Total energies (E) and orbital eigenvalues ($\epsilon$) of the 
  ground and the three lowest excited states of the hydrogen atom computed using 
  DO-MOM with PBE and SIC-PBE and experimental values of the ionization (I)
  energy\cite{nist_hydrogen}. The values in parenthesis represent the differences  
  with respect to the experimental energies. All values are in eV.}
  \label{tbl:Tbl3}
  \arrayrulewidth=1.0pt
  \renewcommand{\arraystretch}{1.5}
  \begin{tabular}{p{3.0cm}M{2.0cm}M{2.0cm}M{2.0cm}M{2.0cm}M{2.0cm}}
    \rowcolor{gray!20}
                 &  \multicolumn{2}{c}{PBE} & \multicolumn{2}{c}{SIC-PBE} & \multicolumn{1}{c}{Exp.\cite{nist_hydrogen}} \\
    \hhline{*1{>{\arrayrulecolor{gray!20}}-}*5{>{\arrayrulecolor{black}}-}}
    \rowcolor{gray!20}
     Electronic state  &  E & $\epsilon$ &  E & $\epsilon$ & -I  \\
     1s                         & -13.60(0.00)     &   -7.59(6.01)    &  -13.60(0.00)     &	 -13.60(0.00) &   -13.60 \\
     2s                         & -3.70(-0.30)     &   -2.23(1.17)    &  -3.40(0.00)     &	 -3.40(0.00) &  -3.40 \\
     2p                        & -3.81(-0.41)     &   -1.91(1.49)    &  -3.40(0.00)     &	 -3.40(0.00) &   -3.40  \\
     3s                        & -1.73(-0.22)     &   -1.13(0.38)    &  -1.50(0.01)     &	 -1.50(0.01) &   -1.51	  \\
  \end{tabular} \\
\end{table}
The PBE functional displays a well-known systematic
underestimation of the energy of the excited states
(linear-response TDDFT with PBE predicts no bound Rydberg states for the hydrogen atom)
\cite{Cheng2008}. The inability of excited-state DFT with KS semi-local functionals 
to 
describe Rydberg series of atoms has been traced back to the fact
that the
long-range form of the effective potential is incorrect (see, for example, reference\cite{Cheng2008}).

The SIE of a one-electron system cancels exactly for the SIC-PBE functional. 
As a result, the SIC-PBE energy values are accurate for the basis set used. 
Furthermore, for a given state the eigenvalue of the occupied 
orbital coincides with the total energy and is independent of the
occupation number, i.e. for a one-electron system the SIC functional restores the derivative discontinuity  
that is missing in the approximate functional\cite{Perdew1981}. 

\subsubsection{Dihydrogen Molecule}
Gill \emph{et al.}\cite{Barca2018} have recently reported SCF-MOM calculations 
of the $^1\Sigma_g^+\added[id=Other]{(1\sigma_g^2 \rightarrow 1\sigma_u^2)}$ doubly excited state of dihydrogen using xc fuctionals 
for several choices of the fraction
of exact exchange. Their results show that GGA and hydrid functionals with small
fraction of HF exchange severely underestimate the excitation energy because the SIE 
in the excited state is significantly larger than in the ground state. The DO-MOM PBE 
calculation of the $^1\Sigma_g^+\added[id=Other]{(1\sigma_g^2 \rightarrow 1\sigma_u^2)}$ state is in line with this observation. The PBE excitation energy
is 27.25 eV, with a deviation of 1.50 eV from the full configuration interaction (CI) result of reference\cite{Barca2018} (28.75 eV). 
The one-electron SIE calculated according to eq. \ref{eq:SIC1} using the density and the orbitals 
converged with PBE
is $\sim$-1.69 eV, compared to an SIE of $\sim$-0.10 eV for the ground state. 
Therefore, most of the error in the excitation energy comes from an imbalance in the SIEs.
If the self-interaction correction is applied non-variationally, the resulting excitation energy 
is equal to 28.83 eV, which is closer to the full CI result. Further improvement is 
obtained with the fully variational SIC-PBE calculations giving an excitation energy 
of 28.79 eV, only 0.04 eV larger than the full CI energy. The remaining error is due 
to the approximate treatment of correlation and to the use of different basis sets
in the DO-MOM SIC-PBE and full CI calculations. 

These results illustrate how self-interaction correction 
in variational DFT calculations of excited states can be an effective route to correct the unbalanced 
SIE between ground and excited states in calculations based on semi-local functionals.

\section{Concluding Remarks}\label{conclusion}
DO has long been known to be a robust and computationally competitive 
alternative to SCF in ground-state calculations\cite{Baarman2011, VandeVondele2003, Voorhis2002}.
Calculations using single-determinant excited-state DFT and DO have been limited to minimization 
of the squared norm of the gradient, while DO of saddle points has 
been considered to be too difficult, due to the need of a better approximation to the Hessian
and the risk of variational collapse.
Here, a DO method is presented that overcomes these challenges by: 
(1) employing a newly developed limited-memory formulation of quasi-Newton SR1 
inverse Hessian update \added[id=Other]{(L-SR1)} that is able to build a better approximation to the Hessian
for saddle-point searches than the \added[id=Other]{L-BFGS} update commonly employed in minimization, 
and
(2) avoiding variational collapse by using MOM constraints.
Since only one gradient evaluation is required at each step, the 
computational cost is the same as for ground-state calculations.
We further note that even if DO-MOM has been presented here in the context of 
excited-state DFT, it can be applied with any other method where the objective is to 
optimize a set of orthogonal orbitals, provided that the appropriate form of the ${\bf L}$ 
matrix is used in the expression of the gradient, eq. \ref{eq:MOMDO12}.

We find that DO-MOM in combination with a localized basis set representation of 
the orbitals outperforms the conventional SCF-MOM approach 
both in terms of robustness and speed of convergence for a benchmark set of 
\added[id=Rev2]{89} excited states. 
The best performance is obtained with the
\added[id=Other]{L-SR1} algorithm when using a memory of 20 iterations.
Furthermore, tests on challenging charge-transfer excited states of nitrobenzene show 
that \added[id=Other]{L-SR1} is more robust than L-BFGS for saddle-point optimization, 
being less dependent on the preconditioner. Therefore, DO-MOM with 
\added[id=Other]{L-SR1} is a promising method for excited-state 
calculations of large systems, where diagonalization of the Hamiltonian matrix needed 
to compute the preconditioner is prohibitive.
These tests were limited to valence and Rydberg excited states of small and medium size 
molecules. 
In future work these tests will be extended to include larger molecules and long-range charge-transfer 
states.

DO-MOM is able to converge single-determinant excited states close to conical intersections, which often 
require fractional occupations in SCF approaches, as demonstrated here for the first two 
singlet excited states of carbon monoxide. This makes it possible to assess more rigorously the applicability
of single-determinant density functional methods for modelling conical intersections as compared to methods 
that explicitly take into account static correlation effects. Crucially, such benchmarks are currently 
lacking despite the fact that excited-state DFT has been proposed in the context of nonadiabatic dynamics 
simulations\added[id=Other]{\cite{Malis2020, Pradhan2018, Maurer2011}}. Formally, the 
single-determinant approximation is a clear limitation of excited-state DFT. 
Multiconfigurational effects can be taken into account within, for example, ensemble DFT 
\cite{Theophilou1979}. Extending the applicability of DO-MOM requires 
handling the simultaneous optimization of the orbitals and the occupation numbers\cite{Nygaard2013}.

Finally, our implementation of DO-MOM can be used with non-unitary invariant functionals, 
such as SIC functionals. \added[id=Rev2]{As pointed out earlier\cite{Zhao2019, Gudmundsdottir2013}, 
the accuracy of excitation energies obtained with semi-local functionals can be affected
by different amounts of SIE in the ground and excited state.
The accurate results obtained from the calculations on the lowest doubly 
excited state of dihydrogen represent a preliminary indication
that SIC functionals can help alleviate this issue. 
However, tests on more complex systems are needed to draw a general conclusion 
on the performance of SIC functionals. Benchmarks on excited states of molecules, 
including Rydberg states, are currently ongoing.
}

\appendix
\section*{Appendix}
\renewcommand{\thesubsection}{\Alph{subsection}}

\renewcommand{\theequation}{\thesubsection.\arabic{equation}}
\numberwithin{equation}{subsection}

\subsection{Exponential Transformation}\label{appendix1}
The spin index is omitted here for simplicity as the exponential transformation does not mix 
orbitals with different spin quantum number.
An initial guess for the optimal orbitals (reference orbitals) is expanded into a linear combination 
of $M$ localised basis functions:
\begin{align}\label{eq:MOMDO1}
\phi_{p}({\bf r}) = \sum_{\mu=1}^{M}C_{\mu p}\chi_{\mu}({\bf r})
\end{align}
The coefficients of this expansion must satisfy the orthonormality constraints:
\begin{align}\label{eq:MOMDO2}
\sum_{\mu\nu}C_{\mu p}^{*}S_{\mu\nu}C_{\nu q} = \delta_{pq}
\end{align}
with:
\begin{align}\label{eq:MOMDO3}
S_{\mu\nu} =\int \chi^{*}_{\mu}({\bf r}) \chi_{\nu}({\bf r}) d{\bf r} 
\end{align}
The optimal orbital coefficients corresponding to an extremum of the energy 
functional can be found through a unitary transformation of the $C_{\mu p}$:
\begin{align}\label{eq:MOMDO4}
O_{\mu k} = \sum_{p=1}^{M}C_{\mu p}\left[ e^{\bm \theta} \right]_{pk} 
\end{align}
The $M \times M$ anti-Hermitian matrix $\bm{\theta}$ contains the parameters 
that describe rotations of the orbitals and is parametrized as:
\begin{align}\label{eq:MOMDO5}
{\bm{\theta}} = \begin{pmatrix} {\bm{\theta}}_{\mr{oo}} & {\bm{\theta}}_{\mr{ov}} \\ -{\bm{\theta}}_{\mr{ov}}^{\dagger} & \bm{0} \end{pmatrix}
\end{align}
where the $N \times N$ block ${\bm{\theta}}_{\mr{oo}}$ contains the parameters 
that describe rotations mixing occupied-occupied (oo) orbitals, while the $N \times (M-N)$ blocks 
${\bm{\theta}}_{\mr{ov}}$ mix occupied-virtual (ov) orbitals. 
The total energy does not depend on rotations among the virtual orbitals and, as a result, the 
virtual-virtual (vv) block is set to zero.
Since $\bm{\theta}$ is anti-Hermitian, the total number of free parameters is $N(2M-N)$. 
For KS functionals, the energy is invariant with respect to unitary transformation of equally 
occupied orbitals and, therefore, the ${\bm{\theta}}_{\mr{oo}}$ block can be set to zero 
without loss of generality~\cite{Hutter1994}. In this case, the number of degrees of freedom is 
reduced to $2N(M-N)$ and the matrix exponential can be calculated using the equation given by 
Hutter \emph{et al.}\cite{Hutter1994}. For SIC functionals, ${\bm{\theta}}_{\mr{oo}}$ cannot be 
set to zero~\cite{Lehtola2016}. In this case, the scaling and squaring algorithm 
of Al-Mohy and Higham\cite{AlMohy2009a} as implemented in the SciPy library\cite{2020SciPy} 
is used to evaluate the matrix exponential.

In order to carry out the optimization efficiently, using a quasi-Newton method, or any other 
gradient-based algorithm, the gradient of the energy with respect to the $\{\theta_{ij}\}$ 
rotation parameters is needed: 
\begin{align}\label{eq:MOMDO12}
\frac{\partial E}{\partial \theta^{*}_{ij}} = \frac{2 - \delta_{ij}}{2} \left[ \int_0^1 e^{t {\bm{\theta}}} {\bf L} e^{-t {\bm{\theta}}}  dt \right]_{ij}
\end{align} 
where the matrix {\bf L} has elements:
\begin{align}\label{eq:MOMDO_LSIC}
L_{lk} =  (f_{l} - f_{k}) H_{lk}
 + f_{k}V_{lk} - f_{l}V_{kl}^{*}
\end{align}
In eq. \ref{eq:MOMDO_LSIC}, the $H_{lk}$ are the elements of the Hamiltonian matrix 
in the basis of optimal orbitals:
\begin{align}\label{eq:MOMDO14}
& H_{lk} = \sum_{\mu\nu} O^*_{\mu l} H_{\mu\nu} O_{\nu k},
& H_{\mu\nu} = \int \chi^{*}_{\mu}({\bf r}) {\bf H}_{\mr{KS}} \chi_{\nu}({\bf r}) d{\bf r} 
\end{align}
while the $V_{lk}$ are the elements of orbital-density dependent potentials due to SIC:
\begin{align}\label{eq:MOMDO_ODP}
& V_{lk} = \sum_{\mu\nu} O^*_{\mu l} V^{k}_{\mu\nu} O_{\nu k},
& V_{\mu\nu}^{k}  = \int \chi^{*}_{\mu}({\bf r}) {\bf V}_{k} \chi_{\nu}({\bf r}) d{\bf r} 
\end{align}
For KS functionals, the $V_{lk}$ become zero.

The integral in eq. \ref{eq:MOMDO12} can be expanded in a series:
\begin{align}\label{eq:MOMDO15}
\int_0^1 e^{t {\bm{\theta}}} {\bf L} e^{-t {\bm{\theta}}}  dt = {\bf L} + \frac{1}{2!} \left[ {\bm{\theta}}, {\bf L}\right] 
                                                                                                    + \frac{1}{3!} \left[ {\bm{\theta}} ,\left[ {\bm{\theta}}, {\bf L} \right]\right] 
                                                                                                    + \ldots
\end{align}
When the norm of the matrix $\bm{\theta}$ is small (${\| \bm{\theta} \|}  \ll 1$), the energy gradient 
can be estimated accurately using only the first term of this series. 
During the optimization, the coefficients of the reference orbitals are updated with
those of the optimal or canonical orbitals at regular step intervals and, in addition, each time the MOM
(see next section) changes the orbital occupations. At every update, the ${\bm{\theta}}$ matrix 
is reset to zero; therefore, these updates avoid ${\| \bm{\theta} \|}$ becoming too large, thus allowing 
to use only the first term of the series in eq. \ref{eq:MOMDO15} to estimate the gradient.

\subsection{Limited-Memory Powell Update}\label{appendix2}
The Powell inverse Hessian update formula in compact form is\cite{SunYuan}:
\begin{align}\label{eq:Powell2}
{\bf B}_{\mr P}^{(k+1)} = {\bf B}^{(k)} + {\bf j}^{(k)}{\bf u}^{T(k)} 
                                          + {\bf u}^{(k)}\left[ {\bf j}^{T(k)} 
                                          - \left( {\bf y}^{T(k)} {\bf j}^{(k)} \right) {\bf u}^{T(k)} \right]
\end{align}
where:
\begin{align}\label{eq:QuasiNewton4}
{\bf u}^{(k)} = \frac{{\bf y}^{(k)}}{{\bf y}^{T(k)}{\bf y}^{(k)}}
\end{align}
and ${\bf j}^{(k)}$ and ${\bf y}^{(k)}$ are defined as in eqs. \ref{eq:QuasiNewton3} 
and \ref{eq:QuasiNewton2}, respectively.
The product ${\bf B}_{\mr P}^{(k)}{\bf v}^{(k)}$, where ${\bf v}^{(k)}$ is a vector,
can be computed using the following recursive formula:
\begin{align}\label{eq:Powell3}
{\bf B}_{\mr P}^{(k)}{\bf v}^{(k)} = &{\bf B}_0^{(k)}{\bf v}^{(k)} 
                                                         + \sum_{i=k-m}^{k-1} {\bf j}^{(i)}{\bf u}^{T(i)}{\bf v}^{(k)}  \nonumber \\
                                                         &+ \sum_{i=k-m}^{k-1}  \left\lbrace {\bf u}^{(i)}\left[ {\bf j}^{T(i)}{\bf v}^{(k)}
                                                         - \left( {\bf y}^{T(i)} {\bf j}^{(i)} \right) {\bf u}^{T(i)}{\bf v}^{(k)} \right] \right\rbrace
\end{align}
The L-Powell algorithm is obtained by replacing the use of eq. \ref{eq:SR3}
with eq. \ref{eq:Powell3} in Algorithm 1, which requires storing the vector
${\bf u}^{(k)}$ in addition to ${\bf j}^{(k)}$ at each step (see also Algorithm 1 
in reference \cite{LeviIvanov2020}).

\subsection{Maximum Overlap Method}\label{appendix3}
The MOM method is used to ensure that the character of the occupied optimal orbitals 
is consistent with the initial guess and to choose the occupation numbers of the canonical 
orbitals whenever they are needed, e.g. when updating the preconditioner
according to eq. \ref{eq:PREC1}. The coefficients of the reference orbitals for the MOM, 
which are used to compute the overlaps with the orbitals at a given step, are chosen as 
the coefficients $C_{\mu p}$ of the orbitals of the initial guess, and are fixed. 
Accordingly, the overlap matrix at step $k$ has elements:
\begin{align}\label{eq:MOM1}
\Omega^{(k)}_{pl} = \sum_{\nu\mu}C^{*}_{p \mu}S_{\mu \nu}O^{(k)}_{\nu l}
\end{align}
where $S_{\mu \nu}$ is defined according to eq. \ref{eq:MOMDO3}. 
The occupied orbitals are chosen as those with the largest projections 
onto the occupied subspace of the initial guess orbitals:
\begin{align}\label{eq:MOM2}
\omega^{(k)}_{l} =\left[ \sum_{p=1}^{N} \left( \Omega^{(k)}_{pl} \right)^2 \right]^{\frac{1}{2}}
\end{align}
If the MOM detects a change of the character of the occupied optimal orbitals, the reference
orbitals for the DO are updated. Analogous expressions are used to obtain the occupation numbers 
of the canonical orbitals when a Hamiltonian diagonalization is performed. 

\begin{acknowledgement}
This work was funded by the Icelandic Research Fund (grant number 196070) 
and the University of Iceland Research Fund. AVI is supported by a doctoral fellowship 
from the University of Iceland. The authors thank Elvar {\"{O}}. Jónsson, Asmus O. Dohn 
and Myneni Hemanadhan for useful discussions.
\end{acknowledgement}

\begin{suppinfo}
Basis set convergence tests, data of the convergence tests, additional DO-MOM calculations
on nitrobenzene, additional information on the SCF-MOM calculations of the excited-state 
potential energy curves of carbon monoxide.
\end{suppinfo}

\clearpage
\bibliography{references}

\end{document}